\documentclass[11pt]{article}
\usepackage{amssymb}
\usepackage{amsmath}
\usepackage{amstext}
\usepackage{graphicx,epsfig}
\usepackage{epsfig}
\usepackage{verbatim} 
\usepackage{empheq,fancybox}
\usepackage{color}
\usepackage{ulem}
\usepackage{enumitem}
\usepackage{subfigure}
\usepackage{bbm}
\usepackage{parskip}
\usepackage{dsfont}
\usepackage[numbers,sort&compress]{natbib}
\usepackage{bm}
\usepackage[dvipsnames]{xcolor}
\usepackage{multirow}
\usepackage{cancel}

\usepackage{hyperref} 
\usepackage[all]{hypcap} 
\hypersetup{
    colorlinks=true,       
    linkcolor=red,          
    citecolor=blue,        
    filecolor=magenta,      
    urlcolor=blue           
}

\usepackage{myterms}

\newcommand{\CME}{\text{\sc \fontfamily{ptm}\selectfont cme}}
\newcommand{\cW}{c_{W}}
\newcommand{\sW}{s_{W}}
\newcommand{\ABC}{\text{\sc abc}}
\newcommand{\UDW}{\text{\sc udw}}
\newcommand{\NEW}{\nu\text{\sc ew}}
\newcommand{\hhW}{{\rm hh}\text{\sc w}}
\newcommand{\Uu}{\text{\sc u}{\rm u}}
\newcommand{\Dd}{\text{\sc d}{\rm d}}
\newcommand{\Ee}{\text{\sc e}{\rm e}}
\newcommand{\hW}{{\rm h}\text{\sc w}}

\newcommand{\h}{{\rm h}}
\newcommand{\Dhu}{\text{\sc d}{\rm hu}}
\newcommand{\Uhu}{\text{\sc u}{\rm hu}}
\newcommand{\Uhd}{\text{\sc u}{\rm hd}}
\newcommand{\Dhd}{\text{\sc d}{\rm hd}}
\newcommand{\Nhe}{\nu{\rm he}}
\newcommand{\Ehe}{\text{\sc e}{\rm he}}

\newcommand{\etoe}{{\rm flip}}
\newcommand{\htoee}{h\leftrightarrow ee}
\newcommand{\CS}{\text{\sc cs}}
\renewcommand{\Gauss}{\ \mathrm{G}}

\numberwithin{equation}{section}

\begin{document}
%
~
\vspace{1truecm}
\renewcommand{\thefootnote}{\fnsymbol{footnote}}
\begin{center}
{\huge \bf{Baryogenesis from \\ \vspace{0.2cm} Decaying Magnetic Helicity}}
\end{center} 

\vspace{1truecm}
\thispagestyle{empty}
\centerline{\Large Kohei Kamada${}^{\rm a}$\footnote{\tt kohei.kamada@asu.edu} and Andrew J. Long${}^{\rm b}$\footnote{\tt andrewjlong@kicp.uchicago.edu}}
\vspace{.7cm}

\centerline{\it ${}^{\rm a}$School of Earth and Space Exploration, Arizona State University,
Tempe, AZ 85287, USA.}

\vspace{.2cm}
\centerline{\it$^{\rm b}$Kavli Institute for Cosmological Physics, University of Chicago, Chicago, Illinois 60637, USA}

\vspace{.5cm}
\begin{abstract}
\vspace{.03cm}
\noindent

As a result of the Standard Model chiral anomalies, baryon number is violated in the early universe in the presence of a hypermagnetic field with varying helicity.  
We investigate whether the matter / anti-matter asymmetry of the universe can be created from the decaying helicity of a primordial (hyper)magnetic field before and after the electroweak phase transition.  
In this model, baryogenesis occurs without $(B-L)$-violation, since the $(B+L)$ asymmetry generated by the hypermagnetic field counteracts the washout by electroweak sphalerons.  
At the electroweak crossover, the hypermagnetic field becomes an electromagnetic field, which does not source $(B+L)$.  
Although the sphalerons remain in equilibrium for a time, washout is avoided since the decaying magnetic helicity sources chirality.  
The relic baryon asymmetry is fixed when the electroweak sphaleron freezes out. 
Under reasonable assumptions, a baryon asymmetry of $n_B / s \simeq 4 \times 10^{-12}$ can be generated from a maximally helical, right-handed (hyper)magnetic field that has a field strength of $B_0 \simeq 10^{-14} \, {\rm Gauss}$ and coherence length of $\lambda_{0} \simeq 1 \, {\rm pc}$ today.  
Relaxing an assumption that relates $\lambda_0$ to $B_0$, the model predicts $n_B / s \gtrsim 10^{-10}$, which could potentially explain the observed baryon asymmetry of the universe.  

\end{abstract}

\newpage

\begingroup
\hypersetup{linkcolor=black}
\tableofcontents
\endgroup

\renewcommand*{\thefootnote}{\arabic{footnote}}
\setcounter{footnote}{0}

\section{Introduction}\label{sec:Introduction}

Among the various problems facing modern cosmology, the origin of the matter / anti-matter asymmetry of the universe is unique in that no direct experimental input is forthcoming.  
The baryon asymmetry has already been measured -- approximately one baryon for every $10^{10}$ photons -- and unlike the problems of dark matter, dark energy, or the primordial density perturbations, there are no dedicated experimental efforts underway to provide additional empirical knowledge about the baryon asymmetry.  
With this consideration in mind, it is appealing to study models in which the baryon asymmetry is created along with some other cosmological relic, such as the dark matter, a network of topological defects, or the primordial magnetic field.  
The hope is that future measurements of the secondary relic could provide insight into the origin of the baryon asymmetry.  

In fact the prospects are very favorable for future observational probes of primordial magnetic fields (see \rref{Durrer:2013pga} for a review).  
A primordial magnetic field generated in the early universe could persist today in the voids between galaxies and clusters, where it would be largely unprocessed by structure formation.  
In recent years, TeV blazars have emerged as a potentially powerful tool for measuring the intergalactic magnetic field (IGMF).  
A deficit of secondary GeV gamma rays observed in blazar spectra points to the presence of an IGMF with field strength $B_0 \gtrsim 10^{-16} \Gauss$ \cite{Neronov:1900zz,Tavecchio:2010mk}.  
Searches for magnetically broadened cascade halos \cite{Ando:2010rb,2011APh....35..135E,Chen:2014rsa} and parity-odd correlators in diffuse gamma ray data \cite{Tashiro:2013ita,Chen:2014qva} have also suggested the existence of an IGMF, which could be of primordial origin \cite{Vachaspati:2016xji}.  
Expecting that future observations will provide additional evidence for an IGMF, we are motivated to study the implications of a magnetic field in the early universe \cite{Long:2015cza}.  

From a theory perspective, there is a robust connection between baryon number violation and gauge fields through the Standard Model chiral anomalies \cite{Adler:1969gk,Bell:1969ts,Hooft:1976up}.  
Since the $\SU{2}_L$ and $\U{1}_Y$ gauge fields have chiral interactions with the Standard Model fermions, quantum effects lead to the violation of baryon and lepton numbers.  
This violation is expressed by the current conservation equations, 
\begin{align}\label{eq:djB}
	\partial_{\mu} j^{\mu}_{B} =\partial_{\mu} j^{\mu}_{L}= N_{\rm g} \frac{g^2}{16\pi^2} {\rm Tr}\bigl[ W_{\mu \nu} \widetilde{W}^{\mu \nu} \bigr] - N_{\rm g} \frac{g^{\prime 2}}{32\pi^2} Y_{\mu \nu} \widetilde{Y}^{\mu \nu} 
	\com
\end{align}
which can be integrated over a finite time interval to give 
\begin{align}\label{eq:Bdot}
	\Delta Q_{B} =\Delta Q_{L} = N_{\rm g} \Delta N_{\CS} - N_{\rm g} \frac{g^{\prime 2}}{16\pi^2} \Delta \Hcal_{Y} 
	\per
\end{align}
Thus changes in baryon and lepton numbers, $Q_B$ and $Q_L$, are induced by changes in $\SU{2}_L$ Chern-Simons number $N_{\CS}$ and $\U{1}_Y$ hypermagnetic helicity $\Hcal_{Y}$.  
For a coherent magnetic field, helicity $\Hcal = \int \! \ud^3 x {\bm A} \cdot {\bm B}$, quantifies the excess of power in either the left- or right-circular polarization mode.  

Many studies have investigated the connection between a primordial magnetic field (PMF) and the baryon asymmetry of the universe (BAU).  
Broadly speaking, the literature falls into three categories:  PMF-from-BAU \cite{Joyce:1997uy,Long:2013tha,Long:2016uez}, BAU-from-PMF \cite{Giovannini:1997eg,Giovannini:1997gp,Giovannini:1999wv,Giovannini:1999by,Bamba:2006km,Bamba:2007hf,Anber:2015yca,Fujita:2016igl}, and co-evolution \cite{Boyarsky:2011uy,Dvornikov:2011ey,Dvornikov:2012rk,Semikoz:2012ka,Dvornikov:2013bca,Semikoz:2013xkc,Semikoz:2015wsa,Zadeh:2015oqf}.  
As emphasized in \rref{Long:2016uez}, it is generally difficult to produce a very strong primordial magnetic field starting from a small baryon asymmetry at the level of $n_B/s \sim 10^{-10}$.  
On the other hand, \rref{Fujita:2016igl} recently pointed out that it is generally easy to over-predict the baryon asymmetry of the Universe from a pre-existing helical magnetic field\footnote{Following \rref{Fujita:2016igl}, we remain agnostic as to the origin of the magnetic field.  Many compelling models of magnetogenesis are summarized in the review \cite{Durrer:2013pga}, including inflationary magnetogenesis and magnetogenesis from a first order symmetry breaking phase transition.}.  
This paper builds on the work of \rref{Fujita:2016igl} with a more sophisticated model for the evolution of the baryon asymmetry across the electroweak crossover. 

When electroweak symmetry breaking occurs at $T \simeq 160 \GeV$, the primordial hypermagnetic field becomes an electromagnetic field.  
Since the $\U{1}_{\rm em}$ gauge field has vector-like interactions with the Standard Model fermions, it does not source baryon and lepton number.  
(There is no $F_{\mu \nu} \tilde{F}^{\mu \nu}$ term on the right side of \eref{eq:djB}.)  
Previously, \cite{Fujita:2016igl} assumed that the baryon asymmetry freezes out at the electroweak phase transition, since the source for $(B+L)$ is absent  ($\Delta \Hcal_{Y}$ term in \eref{eq:Bdot}).  
However, the electroweak sphaleron ($\Delta N_{\rm cs}$ term in \eref{eq:Bdot}) remains in equilibrium until $T \simeq 130 \GeV$ and threatens to washout the $(B+L)$ asymmetry \cite{Giovannini:1997eg}.  
Therefore proper modeling of the epoch $160 \GeV \gtrsim T \gtrsim 130 \GeV$ is critical to an accurate prediction of the relic baryon asymmetry of the Universe.  

The present study builds on earlier work in the following ways:
\begin{enumerate}
	\item  We include kinetic equations for all of the Standard Model fermion species.  Many previous studies have focused on simply the electron asymmetries.  While the electron asymmetries do play a key role, we find that including the quarks and higher-generation leptons allows us to properly implement the transformation of the hypermagnetic field into an electromagnetic field at the electroweak phase transition.  
	\item  We include the chiral magnetic effect (CME).  As we will see, the CME suppresses growth of the baryon asymmetry for models with a strong magnetic field.  The CME was not taken into account in some previous studies.  
	\item  We focus on models with vanishing $(B-L)$.  In this way, we address the question of whether the observed baryon asymmetry of the universe can arise entirely from the decaying magnetic helicity of a primordial magnetic field ({\it i.e.}, BAU-from-PMF).  
\end{enumerate}

Solving the Standard Model kinetic equations in the presence of a primordial (hyper)magnetic field with decaying helicity, we investigate -- both analytically and numerically --  the evolution of the baryon asymmetry during the critical window $160 \GeV \gtrsim T \gtrsim 130 \GeV$.  
We find that the $(B+L)$ asymmetry is not washed out by the electroweak sphaleron even though the hypermagnetic field has been transformed into an electromagnetic field, which does not source $(B+L)$.  
Whereas the electroweak sphaleron efficiently erases the asymmetry of the {\it left-chiral} fermions, which are charged under the electroweak gauge group $\SU{2}_L$, it does not communicate directly with the {\it right-chiral} fermions.  
Thus, the Yukawa interactions or the chiral magnetic effect is necessary to communicate $(B+L)$-violation to the right-chiral fermions.  
However, a total relaxation of both left- and right-chiral fermion asymmetries to zero is prevented by the decaying electromagnetic helicity.  
Although the electromagnetic field does not source $(B+L)$, because of its vector-like interactions, it does source fermion chirality through the standard Adler-Bell-Jackiw anomaly \cite{Adler:1969gk,Bell:1969ts}, and thereby it avoids a complete washout.  
Ultimately, the relic baryon asymmetry is determined by a balance between the source term from decaying magnetic helicity and the washout due to electroweak sphaleron in association with either the electron Yukawa interaction or the chiral magnetic effect.  

The paper is organized as follows.  
In the next section, we formulate kinetic equations for the various Standard Model particle asymmetries, paying particular attention to the source terms that arise from the chiral anomaly in the presence of a helical hypermagnetic field.  
In Sec.~\ref{sec:Analytic}, we solve the kinetic equations in the equilibrium approximation, which yields an analytic expression for the relic baryon asymmetry in terms of the magnetic field strength and coherence length today.  
In Sec.~\ref{sec:Results}, we solve the kinetic equations numerically, demonstrate the reliability of the analytic approximation, and determine the magnetic field parameters leading to maximal baryon asymmetry. 
Section~\ref{sec:Conclusion} is devoted to the conclusion.  

\section{Kinetic Equations}\label{sec:KinEqns}

The baryon asymmetry is distributed among the various Standard Model quarks in the form of particle / anti-particle asymmetries.  
Interactions between the quarks and other Standard Model particles allow these asymmetries to redistribute and evolve.  
In this section, we develop a set of kinetic equations to keep track of the evolution of the various asymmetries.  

\subsection{General Discussion}\label{sec:general}

A generic kinetic equation takes the form
\begin{align}\label{eq:KinEqn_generic}
	\frac{d n_A}{dt} + 3 H n_A = - S_{\ABC} + S^{\rm bkg} + \cdots 
	\per 
\end{align}
In this example, there are three flavors of particles ($A,B,C$) and anti-particles ($\bar{A}, \bar{B}, \bar{C}$).  
In general, each flavor can carry an asymmetry, {\it i.e.} the asymmetry in $A$ is quantified by $n_A$, which equals the number of $A$ particles less the number of $\bar{A}$ anti-particles per unit volume.  
In \sref{sec:full_kin_eqns} we present the full system of kinetic equations for the Standard Model particle asymmetries.  

We have assumed a homogeneous FRW background spacetime with Hubble parameter $H$.  
The term $3 H n_{A}$ accounts for the dilution of density due to cosmological expansion.  

Interactions among the three flavors give rise to the source term 
\begin{align}
	S_{\ABC} = \Gamma_{\ABC} \frac{\mu_A - \mu_B - \mu_C}{6/T^2}
\end{align}
where $\mu_A$ is the chemical potential of species $A$, and $\Gamma_{\ABC}$ is the charge transport coefficient.  
Reactions contributing to $\Gamma_{\ABC}$ include the decay and inverse decay processes, 
\begin{align}\label{eq:decay_examples}
\begin{array}{c|c|c|c|c}
\text{classification} & \multicolumn{3}{|c|}{\text{reaction}} & \text{transp. coeff.} \\ \hline \hline
\multirow{2}{*}{decay} 
& A \to B + C & \bar{B} \to C + \bar{A} & \bar{C} \to \bar{A} + B & \multirow{4}{*}{$\gamma_{\ABC} = \Gamma_{\ABC}/T$} \\
& \bar{A} \to \bar{B} + \bar{C} & B \to \bar{C} + A & C \to A + \bar{B} & \\ \cline{1-4}
\multirow{2}{*}{inverse decay} 
& B + C \to A & C + \bar{A} \to \bar{B} & \bar{A} + B \to \bar{C} & \\
& \bar{B} + \bar{C} \to \bar{A} & \bar{C} + A \to B & A + \bar{B} \to C & 
\end{array}
\com
\end{align}
as well as scattering processes that involve a fourth particle with vanishing chemical potential.  
In general, some of the reactions will be kinematically forbidden.  
In \sref{sec:transport} we discuss the transport coefficients that appear in the Standard Model kinetic equations due to the Yukawa interactions, Higgs self-interactions, and weak gauge interactions.  

The source term $S^{\rm bkg}$ represents the rate per unit volume at which the asymmetry in $A$ grows due to an external (background) impetus that does not depend on the other particle asymmetries.  
In \sref{sec:source} we discuss a source term in the Standard Model kinetic equations induced by the decaying helicity of a primordial magnetic field.  

It is convenient to express the kinetic equation in an alternate form.  
We relate the charge density and chemical potential in the relativistic approximation via 
\begin{align}\label{eq:n_to_mu}
	n_A \approx \frac{1}{6} k_A \mu_A T^2 
	\per
\end{align}
The statistical factor $k_A$ counts the number of internal degrees of freedom ({\it e.g.}, spin, color, and isospin).  
We identify the $A$-abundance 
\begin{align}\label{eq:eta_to_n}
	\eta_A = n_A/s
\end{align}
where $s$ is the entropy density of the cosmological plasma.  
While the universe expands adiabatically we have $ds/dt = -3H$, and \eref{eq:KinEqn_generic} becomes
\begin{align}
	\frac{d \eta_A}{d t} = - \Gamma_{\ABC} \, \left( \frac{\eta_A}{k_A} - \frac{\eta_B}{k_B} - \frac{\eta_C}{k_C} \right) + \frac{S^{\rm bkg}}{s} + \cdots 
	\per
\end{align}
During radiation domination, the Hubble parameter is $H = T^2 / M_0$ with $M_0 \equiv M_{\rm pl} / \sqrt{\pi^2/90 g_{\ast}}$, and $g_{\ast}$ is the effective number of relativistic species in the early universe.  
The age of the universe satisfies $t=1/(2H)$, and by introducing the dimensionless temporal coordinate $x = T/H$ we write the kinetic equation as\footnote{Here we have assumed that $g_{\ast}$ is static, and $x = {\rm const.}/T$.  In the Standard Model at the temperatures of interest, $T \gtrsim 130 \GeV$, it is a good approximation to treat $g_{\ast} \simeq 106.75$ as static.  More generally, one could express the kinetic equation in terms of conformal time, $\ud \tau = \ud t / a(t)$.}  
\begin{align}
	\frac{d \eta_A}{d x} = - \gamma_{\ABC} \, \left( \frac{\eta_A}{k_A} - \frac{\eta_B}{k_B} - \frac{\eta_C}{k_C} \right) + \Scal^{\rm bkg} + \cdots 
	\per
\end{align}
The dimensionless transport coefficient and source have been written as $\gamma_{\ABC} \equiv \Gamma_{\ABC}/T$ and $\Scal^{\rm bkg} = S^{\rm bkg} / (sT)$.  

\subsection{Standard Model Kinetic Equations}\label{sec:full_kin_eqns}

In the Standard Model we are interested in the evolution of various particle asymmetries.  
In the symmetric phase where the Higgs condensate is vanishing, the relevant fermion degrees of freedom are the chiral fermions.  
For instance, the charge abundance $\eta_{u_L^1}$ quantifies the particle / anti-particle asymmetry between the three colors of left-chiral, up-type, first-generation quarks $u_L^1$, and their CP-conjugate anti-particles, $\bar{u}_L^1$.  
That is to say, color is summed: $\eta_{u_L^1} = \eta_{(u_L^1)_{\rm red}} + \eta_{(u_L^1)_{\rm green}} + \eta_{(u_L^1)_{\rm blue}}$.  
The abundance $\eta_{u_L^1}$ is related to the charge density $n_{u_L^1}$ and chemical potential $\mu_{u_L^1}$ as in \erefs{eq:n_to_mu}{eq:eta_to_n}.  

We enumerate the Standard Model particles as left- and right-chiral up-type quarks $u_{L,R}^i$, left- and right-chiral down-type quarks $d_{L,R}^i$, left- and right-chiral electrons $e_{L,R}^i$, left-chiral neutrinos $\nu_L^i$, charged Higgs boson $\phi^+$, neutral Higgs boson $\phi^0$, and charged weak boson $W^+$.  
The index $i$ runs from $1$ to $N_{\rm g} = 3$ and counts the number of fermion generations.  
For each particle, there is a corresponding anti-particle, which we denote with a bar for the fermions and neutral Higgs ({\it e.g.}, $\bar{u}_L^i$ and $\bar{\phi}^0$) and we denote as $\phi^-$, $W^-$ for the charged Higgs and weak boson.  
The statistical factors $k$ that appear in converting from chemical potential $\mu$ to charge abundance $\eta$ are given by\footnote{In general, $k=1$ per degree of freedom for a chiral fermion, $2$ for a Dirac fermion, $2$ for a complex scalar, and $4$ for a complex vector (two spin states).}
\begin{align}\label{eq:stat_factors}
	k_{u_L^i} \approx 
	k_{d_L^i} \approx 
	k_{u_R^i} \approx 
	k_{d_R^i} \approx N_{\rm c} 
	\ , \ \ 
	k_{\nu_L^i} \approx 
	k_{e_L^i} \approx 
	k_{e_R^i} \approx 1
	\ , \ \ 
	k_{\phi^+} \approx 
	k_{\phi^0} \approx 2
	\ , \ \ \text{and} \quad 
	k_{W^+} \approx 4
\end{align}
where $N_{\rm c} = 3$ is the number of colors.  
The neutral gauge bosons ($Y, W^3$ or $\gamma,Z$ in the Higgs phase) are self-conjugate under CP and do not play any role in the charge transport equations.  
At the temperatures of interest the strong interactions are in thermal equilibrium, and we assume that the universe is color-neutral; we assume that the gluons do not carry a charge asymmetry.    

The Standard Model transport equations used in our analysis are summarized below:  
\begin{subequations}\label{eq:KinEqns}
\begin{align}
	\frac{d \eta_{u_L^i}}{dx} & = 
	- \Scal_{\UDW}^{i} 
	- \sum_{j=1}^{N_{\rm g}} \Bigl( \Scal_{\Uhu}^{ij} + \Scal_{\Uu}^{ij} + \Scal_{\Uhd}^{ij} \Bigr)
	- \Scal_{\rm s,sph}
	- \frac{N_{\rm c}}{2} \Scal_{\rm w,sph}
	\nn & \quad 
	+ \Bigl( N_{\rm c} y_{Q_L}^2 \Scal_{\rm y}^{\rm bkg} + \frac{N_{\rm c}}{2} \Scal_{\rm w}^{\rm bkg} + N_{\rm c} \frac{y_{Q_L}}{2} \Scal_{\rm yw}^{\rm bkg} \Bigr)
	\\
	\frac{d \eta_{d_L^i}}{dx} & = 
	\Scal_{\UDW}^{i} 
	- \sum_{j=1}^{N_{\rm g}} \Bigl( \Scal_{\Dhd}^{ij} + \Scal_{\Dd}^{ij} + \Scal_{\Dhu}^{ij} \Bigr)
	- \Scal_{\rm s,sph}
	- \frac{N_{\rm c}}{2} \Scal_{\rm w,sph}
	\nn & \quad 
	+ \Bigl( N_{\rm c} y_{Q_L}^2 \Scal_{\rm y}^{\rm bkg} + \frac{N_{\rm c}}{2} \Scal_{\rm w}^{\rm bkg} - N_{\rm c} \frac{y_{Q_L}}{2} \Scal_{\rm yw}^{\rm bkg} \Bigr)
	\\
	\frac{d \eta_{\nu_L^i}}{dx} & = 
	- \Scal_{\NEW}^{i}
	- \sum_{j=1}^{N_{\rm g}} \Scal_{\Nhe}^{ij}
	- \frac{1}{2} \Scal_{\rm w,sph} 
	+ \Bigl( y_{L_L}^2 \Scal_{\rm y}^{\rm bkg} + \frac{1}{2} \Scal_{\rm w}^{\rm bkg} + \frac{y_{L_L}}{2} \Scal_{\rm yw}^{\rm bkg}\Bigr)
	\\
	\frac{d \eta_{e_L^i}}{dx} & = 
	\Scal_{\NEW}^{i} 
	- \sum_{j=1}^{N_{\rm g}} \Bigl( \Scal_{\Ehe}^{ij} + \Scal_{\Ee}^{ij} \Bigr)
	- \frac{1}{2} \Scal_{\rm w,sph} 
	+ \Bigl( y_{L_L}^2 \Scal_{\rm y}^{\rm bkg} + \frac{1}{2} \Scal_{\rm w}^{\rm bkg} - \frac{y_{L_L}}{2} \Scal_{\rm yw}^{\rm bkg}\Bigr)
	\\
	\frac{d \eta_{u_R^i}}{dx} & = 
	\sum_{j=1}^{N_{\rm g}} \Bigl( \Scal_{\Uhu}^{ji} + \Scal_{\Uu}^{ji} + \Scal_{\Dhu}^{ji} \Bigr)
	+ \Scal_{\rm s,sph}
	- N_{\rm c} y_{u_R}^2 \Scal_{\rm y}^{\rm bkg} 
	\\
	\frac{d \eta_{d_R^i}}{dx} & = 
	\sum_{j=1}^{N_{\rm g}} \Bigl( \Scal_{\Dhd}^{ji} + \Scal_{\Dd}^{ji} + \Scal_{\Uhd}^{ji} \Bigr)
	+ \Scal_{\rm s,sph} 
	- N_{\rm c} y_{d_R}^2 \Scal_{\rm y}^{\rm bkg} 
	\\
	\frac{d \eta_{e_R^i}}{dx} & = 
	\sum_{j=1}^{N_{\rm g}} \Bigl( \Scal_{\Ehe}^{ji} + \Scal_{\Ee}^{ji} + \Scal_{\Nhe}^{ji} \Bigr)
	- y_{e_R}^2 \Scal_{\rm y}^{\rm bkg} \\
	\frac{d \eta_{\phi^+}}{dx} & = 
	- \Bigl( \Scal_{\hhW} + \Scal_{\hW} \Bigr)
	+ \sum_{i,j=1}^{N_{\rm g}} \Bigl( - \Scal_{\Dhu}^{ij} + \Scal_{\Uhd}^{ij} + \Scal_{\Nhe}^{ij} \Bigr)
	\\
	\frac{d \eta_{\phi^0}}{dx} & = 
	\Scal_{\hhW} 
	- \Scal_{\h}
	+ \sum_{i,j=1}^{N_{\rm g}} \Bigl( - \Scal_{\Uhu}^{ij} + \Scal_{\Dhd}^{ij} + \Scal_{\Ehe}^{ij} \Bigr)
	\\
	\frac{d \eta_{W^+}}{dx} & = 
	\Bigl( \Scal_{\hhW} + \Scal_{\hW} \Bigr)
	+ \sum_{i=1}^{N_{\rm g}} \Bigl( \Scal_{\UDW}^{i} + \Scal_{\NEW}^{i} \Bigr)
	\per
\end{align}
\end{subequations}
The source terms fall into the following categories.  
The source terms, 
\begin{subequations}\label{eq:source_gauge}
\begin{align}
	\Scal_{\UDW}^{i} & \equiv \gamma_{\UDW}^{i} \left( \frac{\eta_{u_L^i}}{k_{u_L^i}} - \frac{\eta_{d_L^i}}{k_{d_L^i}} - \frac{\eta_{W^+}}{k_{W^+}} \right) \com \\
	\Scal_{\NEW}^{i} & \equiv \gamma_{\NEW}^{i} \left( \frac{\eta_{\nu_L^i}}{k_{\nu_L^i}} - \frac{\eta_{e_L^i}}{k_{e_L^i}} - \frac{\eta_{W^+}}{k_{W^+}} \right) \com \\
	\Scal_{\hhW} & \equiv \gamma_{\hhW} \left( \frac{\eta_{\phi^+}}{k_{\phi^+}} - \frac{\eta_{\phi^0}}{k_{\phi^0}} - \frac{\eta_{W^+}}{k_{W^+}} \right)
	\com
\end{align}
\end{subequations}
arise from the weak gauge interactions.  
We estimate the corresponding transport coefficients, $\gamma_{\UDW}^{i}$, $\gamma_{\NEW}^{i}$, and $\gamma_{\hhW}$ in \eref{eq:gamQW_approx}.  
The source terms, 
\begin{subequations}\label{eq:source_yukawa}
\begin{align}
	\Scal_{\Dhu}^{ij} & \equiv \frac{\gamma_{\Dhu}^{ij}}{2} \Bigl( \frac{\eta_{d_L^i}}{k_{d_L^i}} + \frac{\eta_{\phi^+}}{k_{\phi^+}} - \frac{\eta_{u_R^j}}{k_{u_R^j}} \Bigr) 
	\quad , \quad
	\Scal_{\Uhu}^{ij} \equiv \frac{\gamma_{\Uhu}^{ij}}{2} \Bigl( \frac{\eta_{u_L^i}}{k_{u_L^i}} + \frac{\eta_{\phi^0}}{k_{\phi^0}} - \frac{\eta_{u_R^j}}{k_{u_R^j}} \Bigr) \com
	\\
	\Scal_{\Uhd}^{ij} & \equiv \frac{\gamma_{\Uhd}^{ij}}{2} \Bigl( \frac{\eta_{u_L^i}}{k_{u_L^i}} - \frac{\eta_{\phi^+}}{k_{\phi^+}} - \frac{\eta_{d_R^j}}{k_{d_R^j}} \Bigr) 
	\quad , \quad
	\Scal_{\Dhd}^{ij} \equiv \frac{\gamma_{\Dhd}^{ij}}{2} \Bigl( \frac{\eta_{d_L^i}}{k_{d_L^i}} - \frac{\eta_{\phi^0}}{k_{\phi^0}} - \frac{\eta_{d_R^j}}{k_{d_R^j}} \Bigr) \com
	\\
	\Scal_{\Nhe}^{ij} & \equiv \frac{\gamma_{\Nhe}^{ij}}{2} \Bigl( \frac{\eta_{\nu_L^i}}{k_{\nu_L^i}} - \frac{\eta_{\phi^+}}{k_{\phi^+}} - \frac{\eta_{e_R^j}}{k_{e_R^j}} \Bigr) 
	\quad , \quad
	\Scal_{\Ehe}^{ij} \equiv \frac{\gamma_{\Ehe}^{ij}}{2} \Bigl( \frac{\eta_{e_L^i}}{k_{e_L^i}} - \frac{\eta_{\phi^0}}{k_{\phi^0}} - \frac{\eta_{e_R^j}}{k_{e_R^j}} \Bigr)
	\com
\end{align}
\end{subequations}
arise from the Yukawa interactions.  
We estimate the transport coefficients in \eref{eq:gammaYuk_approx}.  
After the electroweak phase transition, the gauge and Yukawa interactions mediate scattering with the Higgs condensate.  
This gives rise to the additional source terms,
\begin{subequations}\label{eq:source_condensate}
\begin{align}
	\Scal_{\Uu}^{ij} & \equiv \gamma_{\Uu}^{ij} \Bigl( \frac{\eta_{u_L^i}}{k_{u_L^i}} - \frac{\eta_{u_R^j}} {k_{u_R^j}} \Bigr) \com \\
	\Scal_{\Dd}^{ij} & \equiv \gamma_{\Dd}^{ij} \Bigl( \frac{\eta_{d_L^i}}{k_{d_L^i}} - \frac{\eta_{d_R^j}}{k_{d_R^j}} \Bigr) \com \\
	\Scal_{\Ee}^{ij} & \equiv \gamma_{\Ee}^{ij} \Bigl( \frac{\eta_{e_L^i}}{k_{e_L^i}} - \frac{\eta_{e_R^j}}{k_{e_R^j}} \Bigr) \com \\
	\Scal_{\hW} & \equiv \gamma_{\hW} \left( \frac{\eta_{\phi^+}}{k_{\phi^+}} - \frac{\eta_{W^+}}{k_{W^+}} \right) \com \\
	\Scal_{\h} & \equiv \gamma_{\h} \frac{\eta_{\phi^0}}{k_{\phi^0}}
	\per
\end{align}
\end{subequations}
We estimate these transport coefficients in \eref{eq:gammaYukM_approx}.  

The remaining source terms are associated with the Standard Model chiral anomalies.  
Thermal fluctuations of the non-Abelian gauge fields lead to the terms 
\begin{subequations}\label{eq:source_sphaleron}
\begin{align}
	\Scal_{\rm s,sph} & \equiv \gamma_{\rm s,sph} \sum_{i=1}^{N_{\rm g}} \left( \frac{\eta_{u_L^i}}{k_{u_L^i}} + \frac{\eta_{d_L^i}}{k_{d_L^i}} - \frac{\eta_{u_R^i}}{k_{u_R^i}} - \frac{\eta_{d_R^i}}{k_{d_R^i}} \right) \com \\
	\Scal_{\rm w,sph} & \equiv \gamma_{\rm w,sph} \sum_{i=1}^{N_{\rm g}} \left( \frac{N_{\rm c}}{2} \frac{\eta_{u_L^i}}{k_{u_L^i}} + \frac{N_{\rm c}}{2} \frac{\eta_{d_L^i}}{k_{d_L^i}} + \frac{1}{2} \frac{\eta_{\nu_L^i}}{k_{\nu_L^i}} + \frac{1}{2} \frac{\eta_{e_L^i}}{k_{e_L^i}}  \right) \label{eq:SUDNE}
	\com
\end{align}
\end{subequations}
which are known as the strong and electroweak sphalerons.  
We extract the transport coefficients from the results of lattice simulations in \erefs{eq:gammaS_numb}{eq:gammaW_numb}.  
In the presence of a background magnetic field, additional source terms are generated:
\begin{subequations}\label{eq:source_background}
\begin{align}
	\Scal_{\rm w}^{\rm bkg} & = \begin{cases}
	0 & \quad , \ T > 162 \GeV \\ 
	\frac{1}{2} \left( - \Scal_{\rm em} + \gamma_{\rm em}^{\CME} \, \eta_{5,{\rm em}} \right) & \quad , \ T < 162 \GeV
	\end{cases} \\
	\Scal_{\rm y}^{\rm bkg} & = \begin{cases}
	- \Scal_{\rm y} + \gamma_{\rm y}^{\CME} \, \eta_{5,Y} & \quad , \ T > 162 \GeV \\ 
	- \Scal_{\rm em} + \gamma_{\rm em}^{\CME} \, \eta_{5,{\rm em}} & \quad , \ T < 162 \GeV
	\end{cases} \\
	\Scal_{\rm yw}^{\rm bkg} & = \begin{cases}
	0 & \quad , \ T > 162 \GeV \\ 
	2 \bigl( - \Scal_{\rm em} + \gamma_{\rm em}^{\CME} \, \eta_{5,{\rm em}} \bigr) & \quad , \ T < 162 \GeV 
	\end{cases} 
	\per
\end{align}
\end{subequations}
Above (below) the temperature $T \simeq 162 \GeV$ the system is in the symmetric (broken) phase, see \eref{eq:vT_empirical}.  
The sources $\Scal_{\rm em}$ and $\Scal_{\rm y}$ arise from decaying magnetic helicity, and we estimate them in \erefs{eq:gammaEM_numb}{eq:gammaY_numb}.  
The transport coefficients $\gamma_{\rm em}^{\CME}$ and $\gamma_{\rm y}^{\CME}$ arise from the chiral magnetic effect, and we estimate them in \erefs{eq:gammaEMCME_numb}{eq:gammaYCME_numb}.  
The charge-weighted chiral abundances are defined by  
\begin{align}
	\eta_{5,Y} 
	& = \sum_{i=1}^{N_{\rm g}} \bigl[ -y_{Q_L}^2 \bigl( \eta_{u_L^i} + \eta_{d_L^i} \bigr) - y_{L_L}^2 \bigl( \eta_{\nu_L^i} + \eta_{e_L^i} \bigr) + y_{u_R}^2 \eta_{u_R^i} + y_{d_R}^2 \eta_{d_R^i} + y_{e_R}^2 \eta_{e_R^i} \bigr] \label{eq:eta5Y_def}
	\\
	\eta_{5,{\rm em}} 
	& = \sum_{i=1}^{N_{\rm g}} \bigl[ -q_{u_L}^2 \eta_{u_L^i} - q_{d_L}^2 \eta_{d_L^i} - q_{\nu_L}^2 \eta_{\nu_L^i} - q_{e_L}^2 \eta_{e_L^i} + q_{u_R}^2 \eta_{u_R^i} + q_{d_R}^2 \eta_{d_R^i} + q_{e_R}^2 \eta_{e_R^i} \bigr] \label{eq:eta5em_def}
	\per
\end{align}
The hypercharges and electromagnetic charges of the Standard Model particles are 
\begin{align}
	& y_{Q_L} = \frac{1}{6} \com y_{L_L} = - \frac{1}{2} \com y_{u_R} = \frac{2}{3} \com y_{d_R} = - \frac{1}{3} \com y_{e_R} = - 1 \com y_{\Phi} = \frac{1}{2} \com y_{W} = 0 \\
	& q_{u_L} = q_{u_R} = \frac{2}{3} \com q_{d_L} = q_{d_R} = - \frac{1}{3} \com q_{e_L} = q_{e_R} = -1 \com q_{\nu_L} = 0 \com q_{\phi^+} = q_{W^+} = 1 \com q_{\phi^0} = 0
	\nonumber
	\per
\end{align}
The number of colors is $N_{\rm c} = 3$ and the number of fermion generations is $N_{\rm g} = 3$.  

Let us also define the abundances for hypercharge, electromagnetic charge, baryon number, and lepton number:
\begin{align}
	\eta_{Y} 
	& = \sum_{i=1}^{N_{\rm g}} \bigl[ y_{Q_L} \bigl( \eta_{u_L^i} + \eta_{d_L^i} \bigr) + y_{L_L} \bigl( \eta_{\nu_L^i} + \eta_{e_L^i} \bigr) + y_{u_R} \eta_{u_R^i} + y_{d_R} \eta_{d_R^i} + y_{e_R} \eta_{e_R^i} 
	\nn & \hspace{1.2cm}
	+ y_{\Phi} \bigl( \eta_{\phi^+} + \eta_{\phi^0} \bigr) + y_{W} \eta_{W^+} \bigr] \label{eq:etaY_def}
	\\
	\eta_{{\rm em}} 
	& = \sum_{i=1}^{N_{\rm g}} \bigl[ q_{u_L} \eta_{u_L^i} + q_{d_L} \eta_{d_L^i} + q_{\nu_L} \eta_{\nu_L^i} + q_{e_L} \eta_{e_L^i} + q_{u_R} \eta_{u_R^i} + q_{d_R} \eta_{d_R^i} + q_{e_R} \eta_{e_R^i} 
	\nn & \hspace{1.2cm}
	+ y_{\phi^+} \eta_{\phi^+} + q_{\phi^0} \eta_{\phi^0} + q_{W} \eta_{W^+} \bigr] \label{eq:etaem_def}
	\\
	\eta_{B} 
	& = \frac{1}{N_{\rm c}} \sum_{i=1}^{N_{\rm g}} \bigl[ \eta_{u_L^i} + \eta_{d_L^i} + \eta_{u_R^i} + \eta_{d_R^i} \bigr] \label{eq:etaB_def}
	\\
	\eta_{L} 
	& = \sum_{i=1}^{N_{\rm g}} \bigl[ \eta_{\nu_L^i} + \eta_{e_L^i} + \eta_{e_R^i} \bigr] \label{eq:etaL_def}
	\per
\end{align}
From the system of kinetic equations in \eref{eq:KinEqns}, one can explicitly verify the conservation laws.  
Both electromagnetic charge and baryon-minus-lepton number are conserved, $d\eta_{\rm em}/dx = d(\eta_B-\eta_L)/dx = 0$.  
Hypercharge  is conserved in the symmetric phase, but violated due to the Higgs condensate through the source terms in \eref{eq:source_condensate}.  
The sum baryon-plus-lepton number is violated by the weak sphaleron in \eref{eq:SUDNE} and the background field terms in \eref{eq:source_background}:  
\begin{align}\label{eq:detaBLdx}
	\frac{d(\eta_B+\eta_L)}{dx} = -6 S_{\rm w,sph} + 6 \Scal_{\rm w}^{\rm bkg} - 3 \Scal_{\rm y}^{\rm bkg}
	\per
\end{align}
However, in the broken phase we have $\Scal_{\rm w} = \Scal_{\rm y}/2$ as per \eref{eq:source_background}, and the background electromagnetic field does not violate $(B+L)$.  

\subsection{Charge Transport}\label{sec:transport}

In this section we estimate the charge transport coefficients arising from the charged-current weak interactions, Yukawa interactions, and Higgs condensate.  

\subsubsection{Charged-Current Weak Interactions}\label{sec:gauge}

The left-chiral fermions and the Higgs bosons participate in the charged-current weak interactions with the $W^+$ boson.  
These contributions to the kinetic equations, \eref{eq:KinEqns}, can be identified by the transport coefficients $\gamma_{\UDW}^{i}$, $\gamma_{\NEW}^{i}$, and $\gamma_{\hhW}$.  
The abundances for right-chiral particles ($u_R^i$, $d_R^i$, and $e_R^i$) do not have source terms associated with the weak interactions.  
In fact, these source terms are suppressed by an additional factor of Yukawa coupling squared, and we neglect them.  

The transport coefficients encode various reactions mediated by the weak interactions.  
Some examples of decay reactions are shown in the following table: 
\begin{align}\label{eq:gauge_decay}
\begin{array}{c|c|c|c|c}
\text{classification} & \multicolumn{3}{|c|}{\text{reaction}} & \text{transp. coeff.} \\ \hline \hline
\multirow{2}{*}{decay} & u_L^i \to W^+ d_L^i & W^- \to d_L^i \bar{u}_L^i & \bar{d}^i \to \bar{u}^i W^+ & \multirow{2}{*}{$\gamma_{\UDW}^{i}$} \\
& \bar{u}_L^i \to W^- \bar{d}_L^i & W^+ \to \bar{d}_L^i u_L^i & d_L^i \to u_L^i W^- & \\ \hline \hline
\multirow{2}{*}{decay} & \nu_L^i \to W^+ e_L^i & W^- \to e_L^i \bar{\nu}_L^i & \bar{e}_L^i \to \bar{\nu}_L^i W^+ & \multirow{2}{*}{$\gamma_{\NEW}^{i}$} \\
& \bar{\nu}_L^i \to W^- \bar{e}_L^i & W^+ \to \bar{e}_L^i \nu_L^i & e_L^i \to \nu_L^i W^- & \\ \hline \hline
\multirow{2}{*}{decay} & \phi^+ \to W^+ \phi^0 & W^- \to \phi^0 \phi^- & \bar{\phi}^0 \to \phi^- W^+ & \multirow{2}{*}{$\gamma_{\hhW}$} \\
& \phi^- \to W^- \bar{\phi}^0 & W^+ \to \bar{\phi}^0 \phi^+ & \phi^0 \to \phi^+ W^- & \\
\end{array}
\com
\end{align}
and the corresponding inverse decay reactions are obtained by reversing the direction of the arrow.  
Two-to-two scattering reactions are formed by including a photon, gluon, or $Z$-boson in the initial state.  
Depending on the spectrum, some of the decay and inverse decay reactions will be kinematically forbidden.  
If all decay and inverse decay reactions are forbidden, the transport coefficient arises from scattering reactions, which are suppressed by an additional factor of coupling squared.  

In \aref{app:transport_coeff} we set up the transport coefficient calculation.  
However, a rigorous evaluation of the transport coefficients is beyond the scope of our work.  
Moreover, our calculation of the relic baryon asymmetry is insensitive to these parameters, since the weak interactions come into equilibrium early.  
Thus, we content ourselves with a rough estimate.  

If the transport coefficient arises primarily from one of the decay / inverse decay reactions, we estimate $\gamma$ from the corresponding decay rate.  
At zero temperature, the decay rate is $\Gamma \sim N_{\rm ch} \alpha_{\rm w} m$ where $N_{\rm ch}$ counts the number of decay channels, $\alpha_{\rm w} = g^2/4\pi$ is the weak fine structure constant, and $m$ is the mass of the decaying particle (assumed to be much larger than the mass of the decay products).  
At finite temperature, we must boost from the rest frame of the particle to the rest frame of the plasma where $E \sim p \sim T \gg m$, and $\Gamma$ is suppressed by an additional factor of $m/T$.  
Thus the dimensionless transport coefficient is parametrically given by $\gamma \sim N_{\rm ch} \alpha_{\rm w} m^2/T^2$.  

To obtain a numerical estimate for $\gamma$ we must know $m/T$.  
At high temperatures, particles in the plasma acquire a mass $m \propto T$ where the coefficient equals the coupling constant for the interaction between the particle of interest and the plasma.  
{\it E.g.} for quarks $m/T \sim g_s$ is set by the strong coupling, and for the W-boson $m/T \sim g$ is set by the weak coupling.  
Thus the factor $m^2/T^2$ depends on the spectrum, since $m$ is the mass of the decaying particle.  
To avoid this detail of the calculation, we write $m^2/T^2 \sim 10^{-2}$, in which coupling constants are generically taken to be $O(10^{-1})$.  
With this approach, we estimate the dimensionless transport coefficients as 
\begin{align}\label{eq:gamQW_approx}
	\gamma_{\UDW}^{i} \sim 10^{-2} N_{\rm c} \alpha_{\rm w}
	\ , \quad
	\gamma_{\NEW}^{i} \sim 10^{-2} \alpha_{\rm w}
	\ , \quad \text{and} \quad
	\gamma_{\hhW} \sim 10^{-2} \alpha_{\rm w}
\end{align}
where $N_{\rm c} = 3$ is the number of colors and $\alpha_{\rm w} \simeq 0.033$ is the weak fine structure constant.  
While these estimates are rough, we have verified that our numerical results are insensitive to this ambiguity in the calculation of $\gamma_{\UDW}^i$, $\gamma_{\NEW}^i$, and $\gamma_{\hhW}$.  
Even increasing or decreasing $\gamma$ by a factor of $100$ compared to \eref{eq:gamQW_approx}, we find a negligible change in the relic baryon asymmetry.  

\subsubsection{Yukawa Interactions}\label{sec:yukawa}

The Yukawa interactions allow left-chiral particles to interact with right-chiral particles via a Higgs boson.  
These contributions to the kinetic equations can be identified by the transport coefficients $\gamma_{\Dhu}^{ij}$, $\gamma_{\Uhu}^{ij}$, $\gamma_{\Uhd}^{ij}$, $\gamma_{\Dhd}^{ij}$, $\gamma_{\Nhe}^{ij}$, and $\gamma_{\Ehe}^{ij}$ in \eref{eq:KinEqns}.  
There is no source term for the weak boson $\eta_{W^+}$, because we neglect scattering processes such as $W^{+} d_L^i \to \bar{\phi}^0 u_{R}^{j}$ that are suppressed by an additional factor of $g^2$.  

Decay reactions contributing to the transport coefficients are shown in the following table,
\begin{align}\label{eq:Yukawa_reactions}
\begin{array}{c|c|c|c|c}
\text{classification} & \multicolumn{3}{|c|}{\text{reaction}} & \text{transp. coeff.} \\ \hline \hline
\multirow{2}{*}{decay} 
& \phi^+ \to \bar{d}_L^i u_R^j & d_L^{i} \to \phi^- u_R^j & \bar{u}_R^j \to \bar{d}_L^i \phi^- & \multirow{2}{*}{$\gamma_{\Dhu}^{ij}$} \\
& \phi^- \to d_L^i \bar{u}_R^j & \bar{d}_L^{i} \to \phi^+ \bar{u}_R^j & u_R^j \to d_L^i \phi^+ & \\ \hline \hline
\multirow{2}{*}{decay} 
& \phi^0 \to \bar{u}_L^i u_R^j & u_L^{i} \to \bar{\phi}^0 u_R^j & \bar{u}_R^j \to \bar{u}_L^i \bar{\phi}^0 & \multirow{2}{*}{$\gamma_{\Uhu}^{ij}$} \\
& \bar{\phi}^0 \to u_L^i \bar{u}_R^j & \bar{u}_L^{i} \to \phi^0 \bar{u}_R^j & u_R^j \to u_L^i \phi^0 & \\ \hline \hline
\multirow{2}{*}{decay} 
& \phi^+ \to u_L^i \bar{d}_R^j & \bar{u}_L^i \to \phi^- \bar{d}_R^j & d_R^j \to u_L^i \phi^- & \multirow{2}{*}{$\gamma_{\Uhd}^{ij}$} \\
& \phi^- \to \bar{u}_L^i d_R^j & u_L^i \to \phi^+ d_R^j & \bar{d}_R^j \to \bar{u}_L^i \phi^+ &  \\ \hline \hline
\multirow{2}{*}{decay} 
& \phi^0 \to d_L^i \bar{d}_R^j & \bar{d}_L^i \to \bar{\phi}^0 \bar{d}_R^j & d_R^j \to d_L^i \bar{\phi}^0 & \multirow{2}{*}{$\gamma_{\Dhd}^{ij}$} \\
& \bar{\phi}^0 \to \bar{d}_L^i d_R^j & d_L^i \to \phi^0 d_R^j & \bar{d}_R^j \to \bar{d}_L^i \phi^0 &  \\ \hline \hline
\multirow{2}{*}{decay} 
& \phi^+ \to \nu_L^i \bar{e}_R^j & \bar{\nu}_L^i \to \phi^- \bar{e}_R^j & e_R^j \to \nu_L^i \phi^- & \multirow{2}{*}{$\gamma_{\Nhe}^{ij}$} \\ 
& \phi^- \to \bar{\nu}_L^i e_R^j & \nu_L^i \to \phi^+ e_R^j & \bar{e}_R^j \to \bar{\nu}_L^i \phi^+ & \\ \hline \hline
\multirow{2}{*}{decay} 
& \phi^0 \to e_L^i \bar{e}_R^j & \bar{e}_L^i \to \bar{\phi}^0 \bar{e}_R^j & e_R^j \to e_L^i \bar{\phi}^0 & \multirow{2}{*}{$\gamma_{\Ehe}^{ij}$} \\ 
& \bar{\phi}^0 \to \bar{e}_L^i e_R^j & e_L^i \to \phi^0 e_R^j & \bar{e}_R^j \to \bar{e}_L^i \phi^0 &  
\end{array}
\com
\end{align}
and inverse decay reactions are obtained by time reversal.  
Various scattering reactions can be formed from combinations of decay and inverse decay reactions or by adding an external gauge boson.  

We estimate the transport coefficients using the same approach as in \sref{sec:gauge}.  
See \aref{app:transport_coeff} for additional details.  
The transport coefficients associated with decay and inverse decay reactions via Yukawa interactions can be estimated as 
\begin{align}\label{eq:gammaYuk_approx}
	\gamma_{\Dhu}^{ij} \approx \gamma_{\Uhu}^{ij} \sim 10^{-2} N_{\rm c} \frac{|y_{u}^{ij}|^2}{8\pi}
	\ , \quad 
	\gamma_{\Uhd}^{ij} \approx \gamma_{\Dhd}^{ij} \sim 10^{-2} N_{\rm c} \frac{|y_{d}^{ij}|^2}{8\pi}
	\ , \quad \text{and} \quad 
	\gamma_{\Nhe}^{ij} \approx \gamma_{\Ehe}^{ij} \sim 10^{-2} \frac{|y_{e}^{ij}|^2}{8\pi}
\end{align}
where $N_{\rm c} = 3$ is the number of colors, and the matrices of Yukawa couplings, $y_{u,d,e}^{ij}$ are given in Appendix~B of \rref{Fujita:2016igl}.  
An additional factor of $1/2$ appears, because the coupling is $y/\sqrt{2}$ instead of $g$.  
As discussed in \sref{sec:gauge}, the factor of $10^{-2}$ estimates the ratio $m(T)^2/T^2$, which is different for each channel but generally of order the coupling squared.  

While our results for the relic baryon asymmetry are insensitive most of these transport coefficients, the first generation electron Yukawa interaction plays a more important role.  
The processes in \eref{eq:Yukawa_reactions} involving leptons will violate right-chiral electron number $\eta_{e_R^i}$, and tend to washout these asymmetries.  
Since the (first-generation) electron has the smallest Yukawa coupling, $y_{e}^{11} \simeq 2.8 \times 10^{-6}$, its erasure is least efficient.  
Thus, the survival of a lepton asymmetry stored in $e_R^1$ depends critically on the transport coefficients $\gamma_{\Nhe}^{11}$ and $\gamma_{\Ehe}^{11}$.  
The dominant contribution to these terms, coming from the decay and inverse decays of charged and neutral Higgs bosons, can be evaluated as (see \aref{app:transport_coeff} and also Refs.~\cite{Campbell:1992jd,Cline:1993vv}), 
\begin{align}\label{eq:gammaYuk_e11}
	\gamma_{\Nhe}^{11} \approx \gamma_{\Ehe}^{11} \approx \gamma_{\htoee}
	\qquad \text{with} \qquad
	\gamma_{\htoee} = f_{\htoee} \left(\frac{6 \ln 2}{\pi^2} \frac{|y_{e}^{11}|^2}{8\pi} \frac{m_{h}^2(T)}{T^2}\right)
	\per
	\end{align}
Here $m_{h}^2(T)$ is the temperature-dependent Higgs mass, given in \eref{eq:Higgs_therm_mass}.  
There are a number of assumptions implicit in this approximation of $\gamma_{\htoee}$, and it is not clear that they will remain justified around the electroweak phase transition.  
More generally, $\gamma_{\htoee}$ has a complicated dependence on the left and right-handed electron masses \cite{Lee:2004we,Chung:2009qs} (also see \aref{app:transport_coeff}).  
We expect the prefactor to be $f_{\htoee} \approx 1$ in the symmetric phase where the approximations are reliable, but to parametrize our ignorance, we will consider $f_{\htoee} \neq 1$ in the broken phase.  

\subsubsection{Interaction with Higgs Condensate}\label{sec:condensate}

During the electroweak phase transition, particles may scatter with the background Higgs condensate.  
These scatterings are mediated by Yukawa interactions, Higgs self-interactions, and electroweak gauge interactions.  
The corresponding contributions to the kinetic equations appear with coefficients $\gamma_{\Uu}^{ij}$, $\gamma_{\Dd}^{ij}$, $\gamma_{\Ee}^{ij}$, $\gamma_{\hW}$, and $\gamma_{\h}$ in the SM kinetic equations, \eref{eq:KinEqns}.  

A few of the reactions contributing to the various transport coefficients are shown in the following table:  
\begin{align}\label{eq:Hcond_reactions}
\begin{array}{c|c|c}
\text{classification} & \text{reaction} & \text{transp. coeff.} \\ \hline
\multirow{3}{*}{spin-flip} 
	& u_{L}^{i} \longleftrightarrow u_{R}^{j} & \gamma_{\Uu}^{ij} \\
	& d_{L}^{i} \longleftrightarrow d_{R}^{j} & \gamma_{\Dd}^{ij} \\
	& e_{L}^{i} \longleftrightarrow e_{R}^{j} & \gamma_{\Ee}^{ij} \\ \hline
\multirow{1}{*}{Goldstone mixing} 
	& \phi^{+} \longleftrightarrow W^{+} & \gamma_{\hW} \\ \hline
\multirow{3}{*}{H self-int.} 
	& \phi^{0} \longleftrightarrow \bar{\phi}^{0} & \multirow{4}{*}{$\gamma_{\h}$} \\
	& \phi^{+} \phi^{-} \longleftrightarrow \phi^{0} & \\
	& \phi^{+} \longleftrightarrow \phi^{+} \phi^{0} & \\ \cline{1-2}
\multirow{1}{*}{Goldstone mixing} 
	& \phi^{0} \longleftrightarrow Z & 
\end{array}
	\per
\end{align}
The Yukawa interactions play a particularly important role, since they will tend to erase a chiral asymmetry, and the baryon asymmetry is carried by fermions.  
The Higgs self-interaction and neutral-current weak interactions violate neutral Higgs number, and will tend to drive $\eta_{\phi^0}$ to zero.  
In symmetric phase, the Higgs-self interaction and neutral-current weak interaction conserve particle number and do not enter the kinetic equations.  

We estimate the transport coefficients as follows.  
The 1-to-1 conversion processes in \eref{eq:Hcond_reactions} occur because the Higgs condensate $v(T)$ induces a mass-mixing parameter for the corresponding particles.  
Then, parametrically the transport coefficient must be proportional to the square of this mass parameter.  
Up to $O(1)$ factors, this is $|y_{u,d,e}^{ij}|^2 v(T)^2$ for the quarks and leptons, $\lambda v(T)^2$ for the Higgs boson, $g^2 v(T)^2$ for the W-boson, and $( g^2 + g^{\prime 2}) v(T)^2$ for the Z-boson.  
Here, we have introduced the Higgs self-coupling $\lambda = M_h^2/(2v^2)$ where $v(0) = v \simeq 246 \GeV$ is the Higgs vacuum expectation value today, and $M_h \simeq 125 \GeV$ is the Higgs boson mass.  
Besides the mass factors, the transport coefficient is also proportional to the particle's thermal width $\Gamma_{\rm therm}$.  
If the particle were on-shell, the 1-to-1 conversion reaction would be forbidden.  
The thermal width depends on specifically which decay channels are open, but provided that some channels are open, $\Gamma_{\rm therm}/T$ will generally be of order coupling-squared.  
As in our earlier estimates ({\it cf.}, \eref{eq:gamQW_approx}) we take $\Gamma_{\rm therm} / T \sim 10^{-2}$.  
Combining these factors give our estimates for the condensate-induced transport coefficients: (see also  \aref{app:transport_coeff})
\begin{subequations}\label{eq:gammaYukM_approx}
\begin{align}
	\gamma_{\Uu,\Dd}^{ij} & \sim 10^{-2} \, N_{\rm c} \frac{|y_{u,d}^{ij}|^2}{\pi^2} \frac{v(T)^2}{T^2} \\
	\gamma_{\Ee}^{ij} & \sim 10^{-2} \, \frac{|y_{e}^{ij}|^2}{\pi^2} \frac{v(T)^2}{T^2} \\
	\gamma_{\hW} & \sim 10^{-2} \, \frac{g^2}{\pi^2} \frac{v(T)^2}{T^2} \\
	\gamma_{\h} & \sim 10^{-2} \, \frac{\lambda + \bigl( g^2 + g^{\prime 2} \bigr)}{\pi^2} \frac{v(T)^2}{T^2} 
	\per
\end{align}
\end{subequations}
Recall that $N_{\rm c} = 3$ is the number of colors, $g = \sqrt{4\pi \alpha_{\rm w}} \simeq 0.65$ and $g^{\prime} = \sqrt{4\pi \alpha_{\rm y}} \simeq 0.35$ are the gauge couplings, $\lambda \simeq 0.13$ is the Higgs self-coupling, and the Yukawa matrices $y_{u,d,e}^{ij}$ appear in Appendix~B of \rref{Fujita:2016igl}.  

As we discussed in \sref{sec:yukawa}, the (first-generation) electron can play an important role in maintaining the lepton asymmetry.  
To assess the effect of $\gamma_{\Ee}^{11}$ on the relic baryon asymmetry, we introduce a dimensionless prefactor $f_{\etoe}$ by writing 
\begin{align}\label{eq:gammaYukM_e11}
	\gamma_{\Ee}^{11} \approx \gamma_{\etoe} 
	\qquad \text{with} \qquad
	\gamma_{\etoe} & = f_{\etoe} \Bigl( 10^{-2} \, \frac{|y_{e}^{11}|^2}{\pi^2} \frac{v(T)^2}{T^2} \Bigr)
	\per
\end{align}
By allowing $f_{\etoe} \neq 1$ we parametrize our ignorance of the detailed calculation.  

\subsection{Chiral Anomaly Source Terms}\label{sec:source}

The remaining source terms in the SM kinetic equations (\ref{eq:KinEqns}) arise from the interplay between gauge fields and fermions via chiral anomalies.  
To illustrate the main point, consider the theory of quantum electrodynamics where the axial vector current $j_{5}^{\mu}$ is not conserved, but rather it satisfies\footnote{We denote the dual tensor with a tilde:  $\tilde{F}^{\mu \nu} = (1/2) \epsilon^{\mu \nu \rho \sigma} F_{\rho \sigma}$ and $\epsilon^{0123}=1$. }
\begin{align}
	\partial_{\mu} j^{\mu}_{5} = 2 m \bigl( i \bar{\psi} \gamma^5 \psi \bigr) - 2 \frac{\alpha}{4\pi} F_{\mu \nu} \tilde{F}^{\mu \nu}
	\per
\end{align}
The first term arises from the electron mass $m$, which explicitly violates the chiral symmetry.  
The second term arises from the chiral anomaly, implying that the symmetry is also violated by quantum effects due to interactions of the electron with the photon.  
(For a review, see \cite{Rubakov:1996vz}.)  
The corresponding kinetic equation for the axial charge density $n_5 = \int \! \ud^3x j_5^0$ can be written as 
\begin{align}\label{eq:KinEqn_n5}
	\frac{dn_5}{dx} = - \gamma_{5} n_5 - 2 \frac{1}{sT} \frac{\alpha}{4\pi} \langle \overline{F_{\mu \nu} \tilde{F}^{\mu \nu}} \rangle
\end{align}
where the first term accounts for chirality-flipping reactions ({\it cf}. \eref{eq:source_condensate}).  
The second term is the thermal expectation value (angled brackets) of the volume average (overline) of the pseudoscalar gauge field operator.  
For an Abelian gauge field, we have\footnote{Without taking the volume average we have $-2{\bm E} \cdot {\bm B} = \frac{d}{dt} \bigl( {\bm A} \cdot {\bm B} \bigr) + {\bm \nabla} \cdot \bigl( \phi {\bm B} + {\bm E} \times {\bm A} \bigr)$, and one can see the gauge invariance.  }
\begin{align}\label{eq:FFdual_to_ABdot}
	\langle \overline{F_{\mu \nu} \tilde{F}^{\mu \nu}} \rangle = - 4 \overline{ {\bm E} \cdot {\bm B}} = 2 \frac{d}{dt} \overline{{\bm A} \cdot {\bm B}}
	\com
\end{align}
where $h = \overline{{\bm A} \cdot {\bm B}}$ is the magnetic helicity density.  
Thus, a changing magnetic helicity sources chiral charge, leading to an imbalance between the number of left-chiral and right-chiral fermions.  
In this section, we calculate source terms of this form for the SM gauge fields.  

Denote the $\SU{3}$, $\SU{2}$, and $\U{1}$ field strength tensors by $G_{\mu \nu} = G_{\mu \nu}^{A} t^{A}$, $W_{\mu \nu} = W_{\mu \nu}^{a} \tau^{a}$, and $Y_{\mu \nu}$, respectively.  
The relevant source terms are written as\footnote{Recall that $F_{\mu \nu} = F_{\mu \nu}^{a} T^{a}$ with $T^{a}$ a generator of the $\SU{N}$ Lie algebra, implying ${\rm Tr}\bigl[ F_{\mu \nu} F_{\rho \sigma}\bigr] = (1/2) F_{\mu \nu}^{a} F_{\rho \sigma}^{a}$} 
\begin{subequations}\label{eq:source_def}
\begin{align}
	\Scal_{\rm s} & \equiv \frac{1}{sT} \frac{\alpha_{\rm s}}{4\pi} \langle \overline{{\rm Tr}\bigl[ G_{\mu \nu} \tilde{G}^{\mu \nu} \bigr] } \rangle \\
	\Scal_{\rm w} & \equiv \frac{1}{sT} \frac{\alpha_{\rm w}}{4\pi} \langle \overline{{\rm Tr}\bigl[ W_{\mu \nu} \tilde{W}^{\mu \nu} \bigr] } \rangle \\
	\Scal_{\rm y} & \equiv \frac{1}{sT} \frac{\alpha_{\rm y}}{4\pi} \langle \overline{ Y_{\mu \nu} \tilde{Y}^{\mu \nu} } \rangle \\
	\Scal_{\rm yw} & \equiv \frac{1}{sT} \frac{gg^{\prime}/4\pi}{4\pi} \Bigl[ \langle \overline{ Y_{\mu \nu} \tilde{W}^{3 \mu \nu} } \rangle + \langle \overline{W_{\mu \nu}^3 \tilde{Y}^{\mu \nu} } \rangle \Bigr] 
	\per
\end{align}
\end{subequations}
These source terms contribute to the SM kinetic equations as follows ({\it cf.} Appendix B of \rref{Long:2013tha})
\begin{subequations}\label{eq:KinEqns_source}
\begin{align}
	\frac{d \eta_{u_L^i}}{dx} & = 
	N_{\rm c} y_{Q_L}^2 \Scal_{\rm y} + \frac{N_{\rm c}}{2} \Scal_{\rm w} + N_{\rm c} \frac{y_{Q_L}}{2} \Scal_{\rm yw} + \Scal_{\rm s} + \cdots \\
	\frac{d \eta_{d_L^i}}{dx} & = 
	N_{\rm c} y_{Q_L}^2 \Scal_{\rm y} + \frac{N_{\rm c}}{2} \Scal_{\rm w} - N_{\rm c} \frac{y_{Q_L}}{2} \Scal_{\rm yw}+ \Scal_{\rm s} + \cdots \\
	\frac{d \eta_{\nu_L^i}}{dx} & = 
	y_{L_L}^2 \Scal_{\rm y} + \frac{1}{2} \Scal_{\rm w} + \frac{y_{L_L}}{2} \Scal_{\rm yw} + \cdots \\
	\frac{d \eta_{e_L^i}}{dx} & = 
	y_{L_L}^2 \Scal_{\rm y} + \frac{1}{2} \Scal_{\rm w} - \frac{y_{L_L}}{2} \Scal_{\rm yw} + \cdots \\
	\frac{d \eta_{u_R^i}}{dx} & = 
	- N_{\rm c} y_{u_R}^2 \Scal_{\rm y} - \Scal_{\rm s} + \cdots \\
	\frac{d \eta_{d_R^i}}{dx} & = 
	- N_{\rm c} y_{d_R}^2 \Scal_{\rm y} - \Scal_{\rm s} + \cdots \\
	\frac{d \eta_{e_R^i}}{dx} & = 
	- y_{e_R}^2 \Scal_{\rm y} + \cdots 
	\com
\end{align}
\end{subequations}
and $d\eta_{\phi^+}/dx = d\eta_{\phi^0}/dx = d\eta_{W^+}/dx = 0$. 
It is convenient to separate the background contribution and the contribution arising from thermal fluctuations:
\begin{align}\label{eq:S_decomp}
	\Scal & = \Scal^{\rm bkg} + \Scal^{\rm fluct} 
	\qquad \text{where} \qquad \left\{
	\begin{array}{l}
	\Scal^{\rm bkg} = \frac{1}{sT} \frac{\alpha}{4\pi} \overline{ \langle F \rangle \langle \tilde{F} \rangle } \\
	\Scal^{\rm fluct} = \frac{1}{sT} \frac{\alpha}{4\pi} \left[ \langle \overline{F \tilde{F}} \rangle - \overline{ \langle F \rangle \langle \tilde{F} \rangle } \right] \\
	\end{array}
	\right.
	\per
\end{align}
Here we set the following source terms to zero: 
\begin{align}\label{eq:SSbkg}
	\Scal_{\rm y}^{\rm fluct} = 0, 
	\qquad 
	\Scal_{\rm yw}^{\rm fluct} = 0, 
	\qquad \text{and} \qquad
	\Scal_{\rm s}^{\rm bkg} = 0, 
\end{align}
since in \rref{Long:2016uez} it was shown that the charge erasure due to hypermagnetic helicity fluctuations is inefficient, and  non-Abelian gauge fields receive magnetic masses and are screened by the plasma \cite{Gross:1980br}, which retards the growth of a coherent field.   
In the following we examine other non-trivial contributions. 

\subsubsection*{Chern-Simons Number Diffusion}\label{sec:CS_diffusion}

In the $\SU{3}_c$ and $\SU{2}_L$ sectors, thermal fluctuations of the non-Abelian gauge fields allow Chern-Simons number to diffuse.  
For a general $\SU{N}$ gauge theory, the Chern-Simons number is 
\begin{align}
	Q(t) & = \int_{0}^{t} \! \ud t^{\prime} \int \! \ud^3 x \, \frac{\alpha}{4\pi} {\rm Tr} \bigl[ F_{\mu \nu} \tilde{F}^{\mu \nu} \bigr]
	\per
\end{align}
The diffusive behavior is expressed by
\begin{align}\label{eq:diffusion}
	\langle Q(t)^2 \rangle & = 2\Gamma t V 
\end{align}
where $\Gamma$ is the diffusion coefficient.  
For a system in thermal equilibrium at temperature $T$, lattice simulations give \cite{Moore:1997im}
\begin{align}
	2\Gamma_{\rm s} / T^4 \simeq (108 \pm 15) \alpha_{\rm s}^5
\end{align}
for QCD and \cite{DOnofrio:2014kta}
\begin{align}
	2\Gamma_{\rm w} / T^4 & \simeq \begin{cases}
	(8.0 \pm 1.3) \times 10^{-7} & \ , \ T \gtrsim 161 \GeV \\
	{\rm exp}\Bigl[ -(147.7 \pm 1.9) + (0.83 \pm 0.01) \frac{T}{\rm GeV} \Bigr] & \ , \ T \lesssim 161 \GeV 
	\end{cases} 
\end{align}
for weak isospin.  
In the latter case, the exponential suppression arises at the electroweak phase transition, because the weak gauge bosons become massive.\footnote{Here we assume that the (helical) magnetic field is too weak to change the behavior of the electroweak crossover \cite{Elmfors:1998wz,Comelli:1999gt}.  Specifically, we assume that the results of lattice studies on $\Gamma_{\rm w}$ and $v(T)$ are unchanged.}  

Due to the chiral anomalies, Chern-Simons number diffusion is accompanied by anomalous charge violation, and there is a corresponding source term in the charge transport equations.  
The source term has been calculated in Refs.~\cite{Khlebnikov:1988sr, Mottola:1990bz}.  
In our notation, this corresponds to the fluctuation part of $\Scal_{\rm s}$ and $\Scal_{\rm w}$, defined above, and we have 
\begin{subequations}
\begin{align}
	\Scal_{\rm s}^{\rm fluct} 
	& = - \frac{\Gamma_{\rm s}}{T} \sum_{i=1}^{N_{\rm g}} \frac{\mu_{u_L^i} + \mu_{d_L^i} - \mu_{u_R^i} - \mu_{d_R^i}}{sT} 
	= - \gamma_{\rm s,sph} \sum_{i=1}^{N_{\rm g}} \left( \frac{\eta_{u_L^i}}{k_{u_L^i}} + \frac{\eta_{d_L^i}}{k_{d_L^i}} - \frac{\eta_{u_R^i}}{k_{u_R^i}} - \frac{\eta_{d_R^i}}{k_{d_R^i}} \right) \\
	\Scal_{\rm w}^{\rm fluct} 
	& = - \frac{\Gamma_{\rm w}}{T} \sum_{i=1}^{N_{\rm g}} \frac{N_{\rm c} \frac{\mu_{u_L^i} + \mu_{d_L^i}}{2} + \frac{\mu_{\nu_L^i} + \mu_{e_L^i}}{2}}{sT} 
	= - \gamma_{\rm w,sph} \sum_{i=1}^{N_{\rm g}} \left( \frac{N_{\rm c}}{2} \frac{\eta_{u_L^i}}{k_{u_L^i}} + \frac{N_{\rm c}}{2} \frac{\eta_{d_L^i}}{k_{d_L^i}} + \frac{1}{2} \frac{\eta_{\nu_L^i}}{k_{\nu_L^i}} + \frac{1}{2} \frac{\eta_{e_L^i}}{k_{e_L^i}}  \right)
	\per
\end{align}
\end{subequations}
In the second equality we have defined the transport coefficients
\begin{subequations}
\begin{align}
	\gamma_{\rm s,sph} 
	& \equiv 6 \frac{\Gamma_{\rm s}}{T^4} 
	\simeq (324 \pm 45) \alpha_{\rm s}^5 \label{eq:gammaS_numb}
	\\
	\gamma_{\rm w,sph} 
	& \equiv 6 \frac{\Gamma_{\rm w}}{T^4} 
	\simeq \begin{cases}
	(24.0 \pm 3.9) \times 10^{-7} & \ , \ T \gtrsim 161 \GeV \\
	{\rm exp}\Bigl[ -(146.6 \pm 1.9) + (0.83 \pm 0.01) \frac{T}{\rm GeV} \Bigr] & \ , \ T \lesssim 161 \GeV 
	\end{cases} \label{eq:gammaW_numb}
\end{align}
\end{subequations}
corresponding to the strong and weak sphalerons.  
The factor of $6$ appears when converting between $\mu$ and $\eta$ with \eref{eq:n_to_mu}.  
The QCD fine structure constant is $\alpha_{\rm s} = g_s^2/4\pi \simeq 0.1184$.  

\subsubsection*{Hypercharge \& Weak Isospin Source Terms}\label{sec:source_and_CME}

The three remaining source terms correspond to background contributions from the hypercharge and weak isospin gauge fields.  
Transcribing from \eref{eq:source_def}, they are written as 
\begin{subequations}\label{eq:three_sources}
\begin{align}
	\Scal_{\rm y}^{\rm bkg} & = \frac{1}{sT} \frac{\alpha_{\rm y}}{4\pi} \ \frac{1}{2} \epsilon^{\mu \nu \rho \sigma} \overline{ \langle Y_{\mu \nu} \rangle \langle Y_{\rho \sigma} \rangle } \\
	\Scal_{\rm w}^{\rm bkg} & = \frac{1}{sT} \frac{1}{2} \frac{\alpha_{\rm w}}{4\pi} \ \frac{1}{2} \epsilon^{\mu \nu \rho \sigma} \overline{ \langle W_{\mu \nu}^{a} \rangle \langle W_{\rho \sigma}^{a} \rangle } \\
	\Scal_{\rm yw}^{\rm bkg} & = \frac{1}{sT} \frac{gg^{\prime}/4\pi}{4\pi} \ \epsilon^{\mu \nu \rho \sigma} \overline{ \langle Y_{\mu \nu} \rangle \langle W_{\rho \sigma}^{3} \rangle } 
	\per
\end{align}
\end{subequations}
It is appropriate to discuss these contributions together, because they become entangled after electroweak symmetry breaking.  
In this section, we first calculate these source terms in the symmetric phase, and then consider the broken phase.  

In the symmetric phase, the non-Abelian iso-magnetic field $W^3$ is screened ({\it cf.} \eref{eq:SSbkg}), and the corresponding source terms vanish
\begin{align}
	\Scal_{\rm w}^{\rm bkg} = 0
	\qquad \text{and} \qquad
	\Scal_{\rm yw}^{\rm bkg} = 0
	\qquad \text{(symmetric phase)}
	\per
\end{align}
On the other hand, the Abelian $\U{1}_Y$ magnetic field is not screened.  
The hypercharge source term is written in terms of the hyperelectric and hypermagnetic fields using
\begin{align}\label{eq:FFdual_to_EB}
	\frac{1}{2} \epsilon^{\mu \nu \rho \sigma} \overline{ \langle Y_{\mu \nu} \rangle \langle Y_{\rho \sigma} \rangle } = - 4 \overline{ {\bm E}_{Y} \cdot {\bm B}_{Y} }
	\per
\end{align}
Now we proceed to evaluate this quantity using the equations of chiral magnetohydrodynamics.  
When a medium with a chiral asymmetry is exposed to a magnetic field there is an induced electric current; this  phenomenon is known as the chiral magnetic effect (CME) \cite{Vilenkin:1980fu} (see also the review \cite{Kharzeev:2013ffa}).  
In the context of hypercharge, the induced hyper-electric current is 
\begin{align}
	{\bm j}_{\CME,Y} = \frac{2}{\pi} \alpha_{\rm y} \mu_{5,Y} {\bm B}_{Y}
	\com
\end{align}
where ${\bm B}_{Y}$ is the hypermagnetic field and the chiral asymmetry was given by \eref{eq:eta5Y_def}.  
Then the total hyperelectric current is written as a sum of of dissipative term (Ohm's law) and the non-dissipative term (CME),
\begin{align}
	{\bm j}_{Y} = \sigma_{Y} \bigl( {\bm E}_{Y} + {\bm v} \times {\bm B}_{Y} \bigr) + {\bm j}_{\CME,Y}
\end{align}
where ${\bm v}$ is the local fluid velocity and $\sigma_{Y}$ is the hyperelectric conductivity.  
At high temperature ($T \gg 100 \GeV$) in the symmetric phase, the conductivity is given by \cite{Arnold:2000dr} 
\begin{align}
	\sigma_{Y} \approx 6^4 \zeta(3)^2 \pi^{-3} \left[ \frac{\pi^2}{8} + \frac{22}{3} \right]^{-1} \left( \frac{T}{g^{\prime2} \ln g^{\prime -1}} \right)
	\simeq 55 T
	\per
\end{align}
The hyperelectric current is related to the hypermagnetic field through Ampere's Law,
\begin{align}
	{\bm \nabla} \times {\bm B}_{Y} = {\bm j}_{Y} + \dot{\bm E}_{Y}
	\per
\end{align}
Solving for the hyper-electric field gives
\begin{align}\label{eq:Efield}
	{\bm E}_{Y} = \frac{1}{\sigma_{Y}} {\bm \nabla} \times {\bm B}_{Y} - \frac{1}{\sigma_{Y}} \dot{\bm E}_{Y} - \frac{1}{\sigma_{Y}} {\bm j}_{\CME,Y}  - {\bm v} \times {\bm B}_{Y} 
	\per
\end{align}
We neglect the displacement current, $\dot{\bm E}$, which is justified in the MHD approximation where $|\dot{\bm E}| / |{\bm \nabla} \times {\bm B}| \sim v /c \ll 1$.
The pseudoscalar product is 
\begin{align}\label{eq:E_dot_B}
	-4 {\bm E}_{Y} \cdot {\bm B}_{Y} =  -\frac{4}{\sigma_{Y}} {\bm B}_{Y} \cdot {\bm \nabla} \times {\bm B}_{Y} + \frac{8 \alpha_{\rm y}}{\pi \sigma_{Y}} \mu_{5,Y} {\bm B}_{Y} \cdot {\bm B}_{Y}
	\per
\end{align}
Here, the advection term ${\bm v} \times {\bm B}_{Y}$ has vanished, since it is perpendicular to $ {\bm B}_{Y}$. 
Then the source term becomes
\begin{align}\label{eq:gamY_def_symph}
	\Scal_{\rm y}^{\rm bkg} = - \Scal_{\rm y} + \gamma_{\rm y}^{\CME} \, \eta_{5,Y}(t) 
	\qquad \text{(symmetric phase)}
\end{align}
where
\begin{align}
	\Scal_{\rm y} & \equiv \frac{\alpha_{\rm y}}{\pi \sigma_{Y} s T} \overline{{\bm B}_{Y} \cdot {\bm \nabla} \times {\bm B}_{Y}} \label{eq:gammaY_def} \\
 	\gamma_{\rm y}^{\CME} & \equiv \frac{12}{\pi^2} \alpha_{\rm y}^2 \frac{\overline{{\bm B}_{Y} \cdot {\bm B}_{Y}}}{\sigma_{Y} T^3} \label{eq:gammaYCME_def} 
	\com
\end{align}
and $\alpha_{\rm y} = g^{\prime 2}/4\pi \simeq 0.0097$ is the hypercharge fine structure constant.  
The source term $\Scal_{\rm y}$ is a pseudoscalar, and proportional to the helicity of the hypermagnetic field.  
In the limit of infinite conductivity, the electric field is shorted out, the helicity is conserved ({\it cf.} \eref{eq:FFdual_to_ABdot}), and the source term vanishes.  
Thus we understand: it is the decay of the magnetic helicity due to ohmic dissipation that leads to the source term \cite{Giovannini:1997gp}.  
In order to provide a numerical estimate for the source term and transport coefficient, we require a model for the evolution of the magnetic field, and we return to this issue in the next section.  

Next we consider the broken phase.  
During the electroweak phase transition, the weak gauge fields become massive, and the hypermagnetic field is transformed into an electromagnetic field.  
It would be interesting to study the dynamical evolution of the gauge fields, Higgs condensate, and plasma degrees of freedom across the phase transition, but that analysis is beyond the scope of our work.  
Instead, we adopt the following simplified model.  
See also \rref{Pavlovic:2016mxq} for a recent study of anomalous magnetohydrodynamics around the electroweak phase transition.  

Without loss of generality, we can perform the field redefinition:  
\begin{align}\label{eq:rotations}
	\left. \begin{array}{l}
	W^{+}_{\mu} = \frac{W_{\mu}^{1} - i W_{\mu}^{2}}{\sqrt{2}} \\
	W^{-}_{\mu} = \frac{W_{\mu}^{1} + i W_{\mu}^{2}}{\sqrt{2}} \\
	Z_{\mu} = \cW \, W_{\mu}^{3} - \sW \, Y_{\mu} \\
	A_{\mu} = \sW \, W_{\mu}^{3} + \cW \, Y_{\mu} 
	\end{array} \right\} \longleftrightarrow \left\{
	\begin{array}{l}
	W^{1}_{\mu} = \frac{W_{\mu}^{+} + W_{\mu}^{-}}{\sqrt{2}} \\
	W^{2}_{\mu} = i \frac{W_{\mu}^{+} - W_{\mu}^{-}}{\sqrt{2}} \\
	W_{\mu}^{3} = \cW \, Z_{\mu} + \sW \, A_{\mu} \\
	Y_{\mu} = -\sW \, Z_{\mu} + \cW \, A_{\mu}
	\end{array} \right.
\end{align}
where $\sW$ and $\cW$ are the sine and cosine of the weak mixing angle; we use $\sW^2 = 0.23$.  
In the new basis, the source terms in \eref{eq:three_sources} are written as 
\begin{align}\label{eq:three_sources_again}
	\Scal_{\rm y}^{\rm bkg} 
	& = \frac{1}{sT} \frac{\alpha_{\rm y}}{4\pi} \ \frac{1}{2} \epsilon^{\mu \nu \rho \sigma} \Bigl( \sW^2 \overline{ \langle Z_{\mu \nu} \rangle \langle Z_{\rho \sigma} \rangle } -2 \sW \cW \overline{ \langle Z_{\mu \nu} \rangle \langle A_{\rho \sigma} \rangle } + \cW^2 \overline{ \langle A_{\mu \nu} \rangle \langle A_{\rho \sigma} \rangle } \Bigr) \nn
	\Scal_{\rm w}^{\rm bkg} 
	& = \frac{1}{sT} \frac{\alpha_{\rm w}}{4\pi} \ \frac{1}{4} \epsilon^{\mu \nu \rho \sigma} \Bigl( \cW^2 \overline{ \langle Z_{\mu \nu} \rangle \langle Z_{\rho \sigma} \rangle } + 2 \sW \cW \overline{ \langle Z_{\mu \nu} \rangle \langle A_{\rho \sigma} \rangle } + \sW^2 \overline{ \langle A_{\mu \nu} \rangle \langle A_{\rho \sigma} \rangle } + \cdots \Bigr) \\
	\Scal_{\rm yw}^{\rm bkg} 
	& = \frac{1}{sT} \frac{gg^{\prime}/4\pi}{4\pi} \ \frac{1}{2} \epsilon^{\mu \nu \rho \sigma} \Bigl( -2 \sW \cW \overline{ \langle Z_{\mu \nu} \rangle \langle Z_{\rho \sigma} \rangle } + 2( \cW^2 - \sW^2) \overline{ \langle Z_{\mu \nu} \rangle \langle A_{\rho \sigma} \rangle } + 2 \sW \cW \overline{ \langle A_{\mu \nu} \rangle \langle A_{\rho \sigma} \rangle } + \cdots \Bigr) \nonumber
	\per
\end{align}
The dots indicate terms that vanish as $W^{\pm} \to 0$, which will not be relevant for the following calculation.  
Each of the tensor products can be written in terms of a corresponding electric and magnetic field product with \eref{eq:FFdual_to_EB}.  

\begin{figure}[t]
\hspace{0pt}
\vspace{-0in}
\begin{center}
\includegraphics[width=0.6\textwidth]{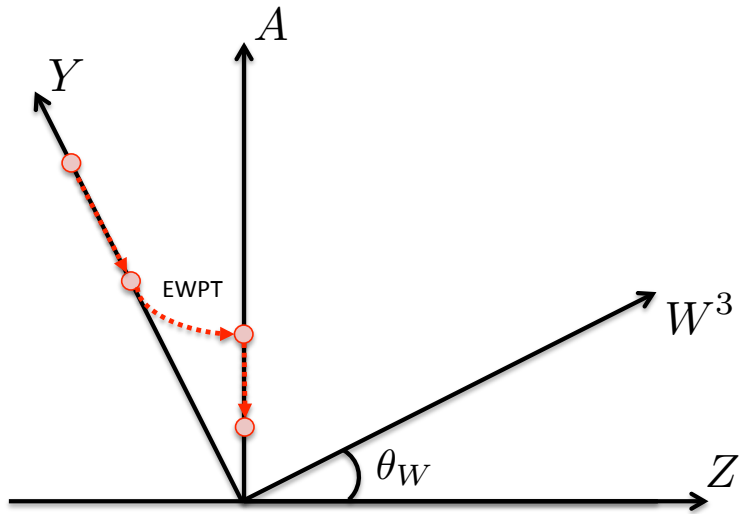}
\caption{
\label{fig:Z-decay_cartoon}
A graphical representation of the evolution from hypermagnetic field to electro-magnetic field during electroweak symmetry breaking.  The Z-component of ${\bm B}_Y$ decays rapidly at the electroweak phase transition / crossover (EWPT).  
}
\end{center}
\end{figure}

Prior to electroweak symmetry breaking, the hypermagnetic field ${\bm B}_Y$ has some components in both ${\bm B}_Z$ and ${\bm B}$ of electromagnetism.  
This is illustrated in \fref{fig:Z-decay_cartoon}.  
The electromagnetic component is calculated from \eref{eq:rotations},
\begin{align}\label{eq:match_B_at_PT}
	{\bm E} = \cW \, {\bm E}_{Y}
	\quad \text{and} \quad
	{\bm B} = \cW \, {\bm B}_{Y}
	\com
\end{align}
noting that $\langle W_{\mu}^{a} \rangle = 0$ in the symmetric phase.  
As the Higgs condensate grows, the W- and Z-field become massive and decay.  
Thus the ${\bm B}_Z$ component of ${\bm B}_Y$ vanishes, leaving only the electromagnetic ${\bm B}$ component.  
We model the decay assuming that the source terms quickly and monotonically decrease to
\begin{align}
	\Scal_{\rm y}^{\rm bkg} 
	= 2 \Scal_{\rm w}^{\rm bkg}
	= \frac{1}{2} \Scal_{\rm yw}^{\rm bkg}
	& = \frac{1}{sT} \frac{\alpha_{\rm em}}{4\pi} (-4) \overline{ {\bm E} \cdot {\bm B} } 
	\com
\end{align}
which is obtained from \eref{eq:three_sources_again} by setting ${\bm E}_Z = {\bm B}_Z = 0$ and using the identities $\alpha_{\rm y} \cW^2 = \alpha_{\rm w} \sW^2 = \alpha_{\rm em}$.  
Then by retracing the calculation that led to \eref{eq:E_dot_B} we obtain
\begin{align}\label{eq:gamY_def_brkph}
	\Scal_{\rm y}^{\rm bkg} = 2 \Scal_{\rm w}^{\rm bkg} = \frac{1}{2} \Scal_{\rm yw}^{\rm bkg}
	= - \Scal_{\rm em} + \gamma_{\rm em}^{\CME} \, \eta_{5,{\rm em}}(t) 
	\qquad \text{(broken phase)}
\end{align}
where
\begin{align}
	\Scal_{\rm em} & \equiv \frac{\alpha_{\rm em}}{\pi \sigma_{\rm em} s T} \overline{{\bm B} \cdot {\bm \nabla} \times {\bm B}} \label{eq:gammaEM_def} \\
 	\gamma_{\rm em}^{\CME} & \equiv \frac{12}{\pi^2} \alpha_{\rm em}^2 \frac{ \overline{{\bm B} \cdot {\bm B}} }{\sigma_{\rm em} T^3} \label{eq:gammaEMCME_def} 
	\com
\end{align}
$\alpha_{\rm em} \simeq 0.0073$ is the electromagnetic fine structure constant, and the electromagnetic conductivity is given by \cite{Arnold:2000dr}
\begin{align}
	\sigma_{\rm em} \simeq (11.9719) \frac{T}{e^2 \ln e^{-1}} \simeq 109 T
	\per
\end{align}
It is important to recognize that the electromagnetic source term $\Scal_{\rm em}$ does not source baryon or lepton number, as we already discussed in \eref{eq:detaBLdx}.  
Thus a decaying helical electromagnetic field itself cannot generate baryon or lepton asymmetry.  

Before closing this section, let us briefly comment on the possible baryon number injection at the time of EWPT. 
Since the hypermagnetic field is assumed to be maximally helical, its ${\bm B}_Z$ component carries some helicity as well.  
When this component decays at the phase transition, there is a large decrease in helicity, and consequently there should be an $O(1)$ change in baryon number due to the chiral anomaly, \eref{eq:Bdot}.  
If the ${\bm B}_Z$ component decays quickly at $T \simeq 160 \GeV$, as we have assumed, then the weak sphaleron is expected to washout the injection of $(B+L)$ and restore the equilibrium solution.  
However, it is interesting to consider that the ${\bm B}_Z$ decay is somehow delayed until after the sphaleron goes out of equilibrium, and some remnant remains in the present Universe, but this is beyond the scope of our study. 

\subsubsection*{Spectrum of Magnetic Field During Inverse Cascade}\label{sec:inverse_cascade}

In order to calculate the source terms and transport coefficients from the previous section, we must evaluate the scalar and pseudoscalar products, $\overline{{\bm B} \cdot {\bm B}}$ and $\overline{{\bm B} \cdot {\bm \nabla} \times {\bm B}}$.  
The scalar product is simply the average magnetic energy density, and the pseudoscalar product is related to the magnetic helicity density.  
At time $t$, suppose that the spectrum peaks at a coherence length scale $\lambda_{B}(t)$ where the peak field strength is $B_p(t)$.  
We assume that the primordial magnetic field is maximally helical, {\it i.e.} the amplitude of one circular polarization mode is much larger than the other.  
Then we estimate 
\begin{align}\label{eq:BB_BdB_approx}
	\overline{{\bm B} \cdot {\bm B}} \approx B_p^2
	\qquad \text{and} \qquad 
	\overline{{\bm B} \cdot {\bm \nabla} \times {\bm B}} \approx \pm 2\pi \frac{B_p^2}{\lambda_B}
	\per
\end{align}
The sign of $\overline{{\bm B} \cdot {\bm \nabla} \times {\bm B}}=0$ is positive for a right-handed magnetic field and negative for a left-handed one \cite{Durrer:2013pga}.  
The additional minus signs in \erefs{eq:gamY_def_symph}{eq:gamY_def_brkph} ensure that a positive (negative) helicity field leads to a negative (positive) source term as it decays ({\it cf.} \eref{eq:FFdual_to_ABdot}).  
Henceforth, we assume a right-handed field.  
Since the kinetic equations \eref{eq:KinEqns} are linear, the solution for a left-handed field is obtained trivially by taking $\eta_f \to - \eta_f$.  
If instead the field were non-helical, the pseudoscalar product would vanish, $\overline{{\bm B} \cdot {\bm \nabla} \times {\bm B}}=0$.  

The evolution of a maximally helical magnetic field in a turbulent plasma has been studied extensively.  
Analytic arguments \cite{FLM:373402, Pouquet:1976zz} and numerical simulations \cite{Kahniashvili:2012uj} reveal that helicity is approximately conserved while power is transported from small scales to larger ones through an inverse cascade.  
In this situation, the spectrum develops with a characteristic scaling law \cite{Campanelli:2007tc}.  
After recombination the plasma is neutral, and the magnetic field evolves adiabatically due to the expansion of the universe.  
Using this scaling relation, the spectrum of the primordial magnetic field in the early universe is expressed in terms of the coherence length and field strength today, $\lambda_{0}$ and $B_{0}$ as \cite{Fujita:2016igl} (see also \aref{app:Bplamb00})\footnote{More generally, the (hyper)magnetic field may evolve approximately adiabatically at first if its initial coherence length is large and field strength is weak, {\it i.e.} $\lambda_B > v_{A} t$ with $v_{A} \propto B_p$ the Alfv\'en velocity \cite{Fujita:2016igl}.  For a causally generated, maximally-helical (hyper)magnetic field, the inverse cascade regime is typically reached before the electroweak crossover \cite{Fujita:2016igl}.  }
\begin{align}
	B_p & \simeq (1 \times 10^{20} \Gauss) \left( \frac{T}{100 \GeV} \right)^{7/3} \left( \frac{B_0}{10^{-14} \Gauss} \right) \mathcal{G}_{B}(T) \label{eq:Bp_numb} 
	\\
	\lambda_B & \simeq (2 \times 10^{-29} \Mpc) \left( \frac{T}{100 \GeV} \right)^{-5/3} \left( \frac{\lambda_0}{1 \pc} \right) \mathcal{G}_{\lambda}(T) \label{eq:lamB_numb}
\end{align}
where $\mathcal{G}_B$ and $\mathcal{G}_{\lambda}$ are $O(1)$ factors that depend on the number of relativistic species.  

When the inverse cascade terminates at the time of recombination, causality considerations restrict the coherence length to be comparable to the largest processed eddy scale $\lambda_B \sim v_A t_{\rm rec}$ where the Alfven velocity $v_A$ depends on the magnetic field strength.  
Thus one obtains the linear relation \cite{Banerjee:2004df}
\begin{align}\label{eq:lam0_to_B0}
	\left( \frac{\lambda_0}{1 \pc} \right) \sim \left( \frac{B_0}{10^{-14} \Gauss} \right)
\end{align}
which is expected to be maintained for a causally generated primordial magnetic field.  
The constant of proportionality is model-dependent \cite{Banerjee:2004df}.  
Comparing with numerical simulations \cite{Kahniashvili:2012uj} we infer that $\lambda_{0}/B_0$ can range from $O(0.1)$ to $O(1)$ in the units above.  

Now we can make numerical estimates of the source terms and CME transport coefficients.  
Evaluating \erefs{eq:gammaEM_def}{eq:gammaEMCME_def} with the inverse cascade scaling relations above, we obtain 
\begin{align}
	\Scal_{\rm em} & \simeq (4 \times 10^{-3}) x^{-4/3} \left( \frac{B_{0}}{10^{-14} \Gauss} \right)^{2} \left( \frac{\lambda_{0}}{1 \pc} \right)^{-1} \label{eq:gammaEM_numb} \\
 	\gamma_{\rm em}^{\CME} & \simeq (1 \times 10^{-2}) x^{-2/3} \left( \frac{B_{0}}{10^{-14} \Gauss} \right)^{2} \label{eq:gammaEMCME_numb} 
	\per
\end{align}
Matching the electromagnetic field to the hypermagnetic field at the time of the phase transition, \eref{eq:match_B_at_PT}, we extrapolate the scaling relation into the symmetric phase.  
Then evaluating \erefs{eq:gammaY_def}{eq:gammaYCME_def} gives 
\begin{align}
	\Scal_{\rm y} & \simeq (1 \times 10^{-2}) x^{-4/3} \left( \frac{B_{0}}{10^{-14} \Gauss} \right)^{2} \left( \frac{\lambda_{0}}{1 \pc} \right)^{-1} \label{eq:gammaY_numb} \\
 	\gamma_{\rm y}^{\CME} & \simeq (4 \times 10^{-2}) x^{-2/3} \left( \frac{B_{0}}{10^{-14} \Gauss} \right)^{2} \label{eq:gammaYCME_numb} 
	\per
\end{align}
If the field were non-helical, we would have $\Scal_{\rm em} = \Scal_{\rm y} = 0$ instead.  

\section{Analytic Equilibrium Solution}\label{sec:Analytic}

Now we search for an equilibrium solution of the kinetic equations, \eref{eq:KinEqns}.  

\subsection{General Considerations} 

The kinetic equations are linear in the various abundances, $\eta(x)$.  
Thus, it is convenient to represent them as a matrix equation
\begin{align}\label{eq:kin_eqns_vector}
	\frac{d}{d x} \vec{\eta} = \mathbb{M} \, \vec{\eta} + \vec{\Scal} 
	\per
\end{align}
Here we identify the vector of charge abundances,
\begin{align}
	\vec{\eta} = \bigl( \, \eta_{u_L^1} \, , \, \eta_{u_L^2} \, , \, \eta_{u_L^3} \, , \, \eta_{d_L^1} \, , \, \cdots \, , \, \eta_{e_R^3} \, , \, \eta_{\phi^+} \, , \, \eta_{\phi^0} \, , \, \eta_{W^+} \, \bigr)^T
	\per
\end{align}
The source vector $\vec{\Scal}$ depends on $\Scal_{\rm y}$ and $\Scal_{\rm em}$, and it arises from the decaying hypermagnetic helicity.
The matrix $\mathbb{M}$ depends on the various transport coefficients ($\gamma$'s), and it is responsible for washout.  

Since both $\vec{\Scal}$ and  $\mathbb{M}$  vary slowly, the system quickly reaches equilibrium where 
\eref{eq:kin_eqns_vector} reduces to a set of algebraic equations 
\begin{align}\label{eq:equilib_matrix_eqn}
	0 = \vec{\Scal} + \mathbb{M} \, \vec{\eta}_{\rm eq}
	\per
\end{align}
Formally, the solution is $\vec{\eta}_{\rm eq} = - \mathbb{M}^{-1} \, \vec{\Scal}$, but in general the matrix $\mathbb{M}$ is singular, and its inverse does not exist.
For each conservation law encoded in the kinetic equations, $\mathbb{M}$ has a vanishing eigenvalue.  
By choosing values for the conserved charges, {\it e.g.} $\eta_{B}-\eta_{L}=0$ and $\eta_{\rm em} = 0$, the equilibrium solution can be obtained.  
One can also consider initial conditions with $\eta_{B} - \eta_{L} \neq 0$ \cite{Harvey:1990qw}, but the case $\eta_{B} - \eta_{L} = 0$ is particularly interesting, because a relic baryon asymmetry may be generated without violating $(B-L)$ and despite the EW sphaleron being in thermal equilibrium.  

\subsection{Simplifying Assumptions} 

To study the full Standard Model, we set the number of generations to three ($N_{\rm g}=3$).  
However, this leads to a set of $7N_{\rm g}+3 = 24$ kinetic equations plus additional conservation laws, which is intractable.  
Instead of solving the full equilibrium equations \eqref{eq:equilib_matrix_eqn}, we impose several approximations to simplify them. 

It is known that without the source term, the baryon and lepton asymmetries are washed out due to the electroweak sphaleron if there is no initial $(B-L)$ asymmetry \cite{Kuzmin:1985mm}.  
However, the electroweak sphaleron cannot complete the washout by itself, because it only communicates with the left-chiral fermions, but the Yukawa interactions and weak interactions must also be in equilibrium \cite{Harvey:1990qw}.  
As a result, the washout does not complete until the first generation electron Yukawa interaction enters chemical equilibrium \cite{Campbell:1992jd}. 
This fact lets us simplify the analysis:  we impose chemical equilibrium for the strong sphaleron, electroweak sphaleron, weak interaction\footnote{The chemical equilibrium of weak interaction  becomes less reliable in the broken phase as the weak gauge bosons become massive and go out of equilibrium.  However, we have verified numerically that the approximation is robust down to temperature of $T \gtrsim 130 \GeV$.  }, Yukawa interactions except for the first generation electron, and we impose $\eta_{B} - \eta_{L} = \eta_{\rm em} = 0$.  
Since the structure of the kinetic equations are different  in the symmetric and broken phases, in the following, we give the analytic equilibrium solutions separately. 

\subsection{Symmetric Phase} 

In the symmetric phase, we impose chemical equilibrium of the strong sphaleron ($\Scal_{\rm s,sph} = 0$), electroweak sphaleron ($\Scal_{\rm w,sph} = 0$), weak interactions ($\Scal_{\UDW}^{i} = \Scal_{\NEW}^{i} = \Scal_{\hhW} = 0$), Yukawa interactions ($\Scal_{\Dhu}^{ij} = \cdots = \Scal_{\Ehe}^{ij} = 0$ except for $\Scal_{\Ehe}^{11}$ and  $\Scal_{\Nhe}^{11}$), and we assume that the conserved charges are vanishing $\eta_{B} - \eta_{L} = \eta_{\rm em} = 0$ ({\it cf.} Eqs.~(\ref{eq:etaem_def}),~(\ref{eq:etaB_def}),~and~(\ref{eq:etaL_def})).  
Since there is no source for $W^+$, we also have $\eta_{W^+} \approx 0$ to a good approximation.  
Since the Higgs condensate vanishes, we do not impose the equilibrium  conditions for these interactions.  

With these equilibrium conditions, we find that the equilibrium values of $\eta_B$ and $\eta_{5,Y}$ are related to the abundances of first generation leptons and Higgs bosons as 
\begin{align}\label{eq:}
\eta_{5,Y,{\rm eq}}=\eta_{e_R^1,{\rm eq}}+\frac{\eta_{\phi^+,{\rm eq}}+\eta_{\phi^0,{\rm eq}}}{4}-\frac{\eta_{\nu_L^1,{\rm eq}}+\eta_{e_L^1,{\rm eq}}}{2} = \frac{79}{22} \eta_{B,{\rm eq}} \per
\end{align}
The equilibrium equation for the first generation right-handed electron is then expressed as 
\begin{equation} \label{eq:BAU_eq_approx_symph}
	0 \approx \Scal_{\rm y} - \frac{79}{22} ( \gamma_{\htoee}  + \gamma_{\rm y}^{\CME} ) \eta_{B,{\rm eq}} 
	\com
\end{equation}
and we reach the equilibrium solution 
\begin{align}\label{eq:BAUeq_symph}
	\eta_{B,{\rm eq}} = \eta_{L,{\rm eq}} 
	\approx \frac{22}{79} \frac{\Scal_{\rm y} }{ \gamma_{\htoee} + \gamma_{\rm y}^{\CME} } 
	\qquad \text{(symmetric phase)}
	\per
\end{align}
The definitions of $\gamma_{\htoee} $, $\Scal_{\rm y}$, and $\gamma_{\rm y}^{\CME}$ can be found in Eqs.~(\ref{eq:gammaYuk_e11}),~(\ref{eq:gammaY_def}),~and~(\ref{eq:gammaYCME_def}).  
Equation~\ref{eq:BAUeq_symph} expresses the competition between the decaying hypermagnetic helicity ($\Scal_{\rm y}$), which tends to grow the baryon asymmetry, and the washout by Yukawa interactions ($\gamma_{\htoee}$) and chiral magnetic effect $(\gamma_{\rm y}^{\CME}$).  
An equation similar to \eref{eq:BAUeq_symph} first appeared in Refs.~\cite{Giovannini:1997eg,Giovannini:1997gp} where the right-chiral electron asymmetry was calculated in the presence of a hypermagnetic field.  

Assuming the hypermagnetic field experience the inverse cascade, we have calculated $\Scal_{\rm y}$ and $\gamma_{\rm y}^{\CME}$ in \erefs{eq:gammaY_numb}{eq:gammaYCME_numb}.  
Combining these expressions gives an analytic approximation to the baryon asymmetry
\begin{align}\label{eq:BAUeq_numb_symph}
	\eta_{B,{\rm eq}} 
	\simeq \bigl( 4 \times 10^{-12} \bigr) \, \frac{B_{14}^2}{\lambda_1} \, \frac{(x/x_{\rm w})^{-4/3} }{0.08 f_{\htoee} \frac{m_h(T)^2}{T^2}+ B_{14}^2 \, (x/x_{\rm w})^{-2/3} } 
	\qquad \text{(for $x < x_{\rm w}$)}
	\per
\end{align}
Here we define $B_{14} \equiv B_{0} / (10^{-14} \Gauss)$ and $\lambda_1 \equiv \lambda_0/(1 \pc)$.  
We write $(x/x_{\rm w}) = (T_{\rm w}/T)$ with $x_{\rm w} \simeq 4.4 \times 10^{15}$ or $T_{\rm w} \simeq 162 \GeV$ corresponding approximately to the time of the electroweak phase transition.  
Recall that $f_{\htoee}$ is a dimensionless prefactor associated with our uncertainty in the calculation of the transport coefficient for Higgs decay and inverse decay.
The above approximation is reliable in the symmetric phase when $x < x_{\rm w}$.  

\subsection{Broken Phase} 

Next we consider the equilibrium solution at temperatures $130 \GeV \lesssim T \lesssim 160 \GeV$ in the broken phase.  
The analysis is similar to the symmetric phase (previous section) with the exceptions that we cannot neglect $\eta_{W^+}$ and we should impose chemical equilibrium for the reactions involving the Higgs condensate ($\Scal_{\Uu}^{ij} = \Scal_{\Dd}^{ij} = \Scal_{\Ee}^{ij} = \Scal_{\hW} = \Scal_{\h} = 0$ except for $\Scal_{\Ee}^{1}$).  
We find that the equilibrium values of $\eta_B$ and $\eta_{5,{\rm em}}$ are related to the chiral asymmetry of electrons, 
\begin{equation}
	(\eta_{e_R^1,{\rm eq}}-\eta_{e_L^1,{\rm eq}}) \approx \eta_{5,{\rm em},{\rm eq}} \approx \frac{37}{11} \eta_{B,{\rm eq}} 
	\per
\end{equation}
Thus, the equilibrium equation for the first generation right-handed electron becomes
\begin{equation}
	0 \approx \Scal_{\rm em}- \frac{37}{11}(  \gamma_{\htoee} + \gamma_{\etoe}+\gamma_{\rm em}^{\CME}) \eta_{B,{\rm eq}} 
	\com
\end{equation}
and the equilibrium solution reduces to 
\begin{align}\label{eq:BAUeq_brkph}
	\eta_{B,{\rm eq}} = \eta_{L,{\rm eq}} 
	\approx \frac{11}{37} \frac{\Scal_{\rm em} }{  \gamma_{\htoee} + \gamma_{\etoe}+ \gamma_{\rm em}^{\CME} } 
	\qquad \text{(broken phase)}
	\per
\end{align}
The source term ($\Scal_{\rm em}$) associated with the electromagnetic field was defined in \eref{eq:gammaEM_def}, and the transport coefficients associated with the electron chirality-flipping reactions ($\gamma_{\htoee}$, $\gamma_{\etoe}$) and the chiral magnetic effect ($\gamma_{\rm em}^{\CME}$) were defined in Eqs.~\eqref{eq:gammaYuk_e11}, \eqref{eq:gammaYukM_e11}, and \eqref{eq:gammaEMCME_def}.   

The important point is that the baryon and lepton asymmetry is generated even though the source term from decaying (electro)magnetic helicity does not violate $(B+L)$-number.  
This is because the electroweak sphaleron itself can only affect the asymmetries of left-chiral fermions, which are charged under $\SU{2}_L$.  
The Yukawa interactions ($\gamma_{\htoee}$, $\gamma_{\etoe}$) or chiral magnetic effect ($\gamma_{\rm em}^{\CME}$) is required to communicate $(B+L)$-violation to the right-chiral fermions.  
However, the electromagnetic field sources chirality ($\Scal_{\rm em}$) preventing a complete equilibration.  
Thus, the baryon asymmetry in \eref{eq:BAUeq_brkph} is understood to result from a competition between the source of fermion chirality and the processes that would wash it out. 

For a maximally helical magnetic field that undergoes the inverse cascade, the source term and CME transport coefficient were calculated in \erefs{eq:gammaEM_numb}{eq:gammaEMCME_numb}.   
Using these expressions, we obtain an approximate solution for the baryon asymmetry, 
\begin{align}\label{eq:BAUeq_numb_brkph}
	\eta_{B,{\rm eq}} 
	\approx (5\times 10^{-12} ) \, \frac{B_{14}^2}{\lambda_{1}} \, \frac{ (x/x_{\rm w})^{-4/3} }{0.4 \, f_{\htoee} \frac{m_h(T)^2}{T^2} + 0.02 \, f_\etoe \frac{v(T)^2}{T^2} + B_{14}^2 \ (x/x_{\rm w})^{-2/3}}
	\qquad \text{(for $x > x_{\rm w}$)}
\end{align}
where the notation is defined below \eref{eq:BAUeq_numb_symph}.  
Recall that the factor $f_{\etoe} \sim 1$ parametrizes our ignorance of the transport coefficient for electron spin flip.  

The approximation in \eref{eq:BAUeq_numb_brkph} is reliable until the electroweak sphaleron freezes out at $T\simeq 135 \GeV$.  
Afterward baryon number is conserved, and $\eta_B$ is constant under adiabatic evolution.  
Then the relic baryon asymmetry today is given by 
\begin{align}\label{eq:BAUeq_numb_today}
	\eta_{B} (t_0) = \eta_{L}  (t_0) = \eta_{B,{\rm eq}}(T\simeq 135 {\rm GeV}) 
	\approx (4\times 10^{-12} )\frac{ B_{14}^2 / \lambda_1}{0.2 f_{\htoee}+ 0.04 f_{\etoe} +B_{14}^2 }
	\per
\end{align}  
In the next section, we verify this approximation against a full numerical solution of the kinetic equations.

\section{Numerical Results and Predictions}\label{sec:Results}

In this section, we investigate numerical solutions of the kinetic equations, \eref{eq:KinEqns}.  
As an initial condition, we assume that all asymmetries are vanishing at an initial temperature $T_{\rm ini}$.  
In effect, this assumes that magnetogenesis occurs rapidly at $T_{\rm ini}$, producing a maximally helical magnetic field without generating any significant particle-number asymmetries.  
The three model parameters are the magnetic field injection temperature $T_{\rm ini}$, the magnetic field strength today $B_0$, and the coherence length today $\lambda_0$.  
Additionally, in estimating the charge transport coefficients, we introduced ``fudge factors'' to parametrize our ignorance of the detailed calculation.  
These two factors affect the rate of Higgs inverse decays ($f_{\htoee}$) and electron spin-flips ($f_{\etoe}$).  

\begin{figure}[t]
\begin{center}
\includegraphics[width=0.49\textwidth]{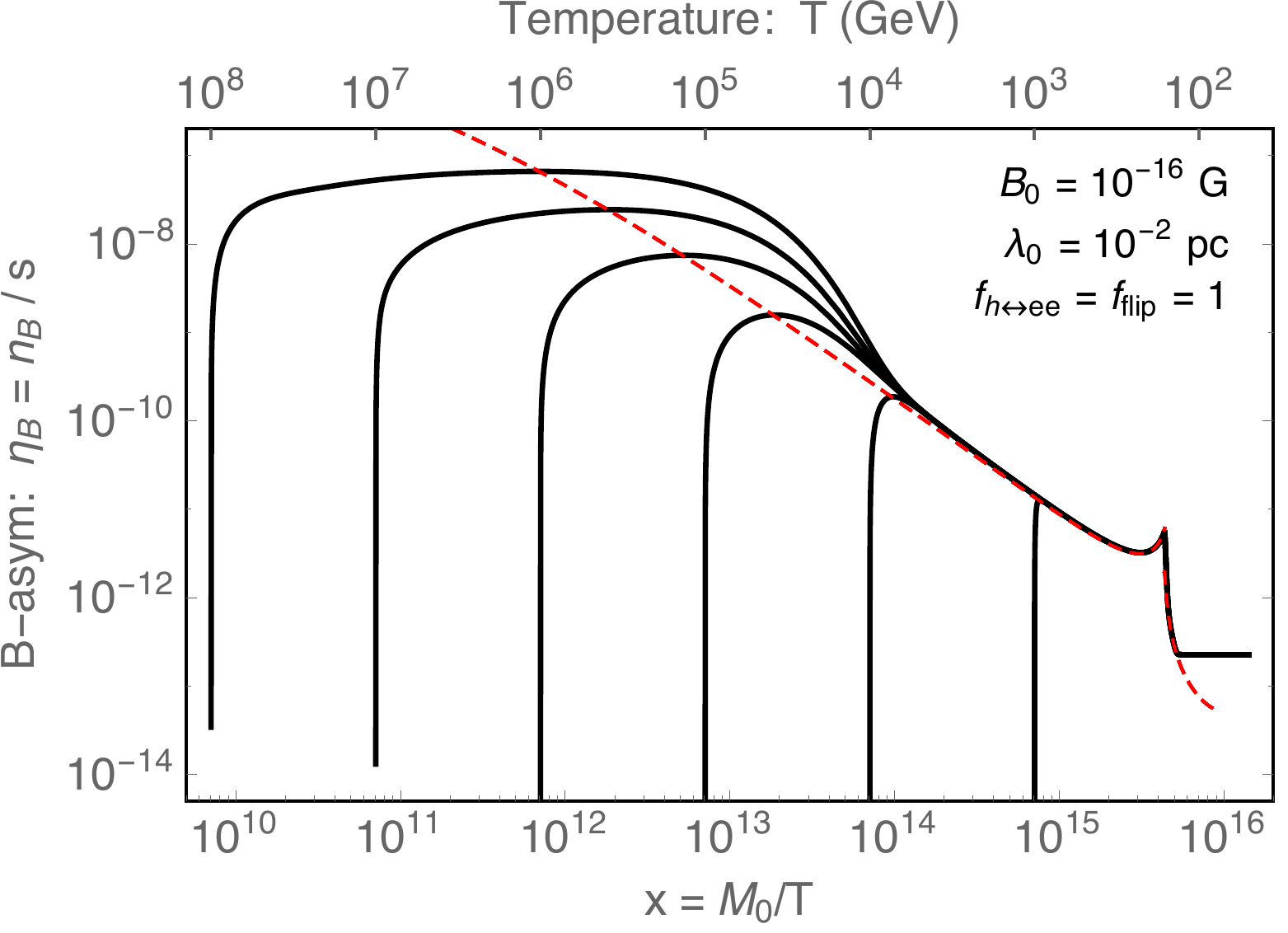} \hfill
\includegraphics[width=0.49\textwidth]{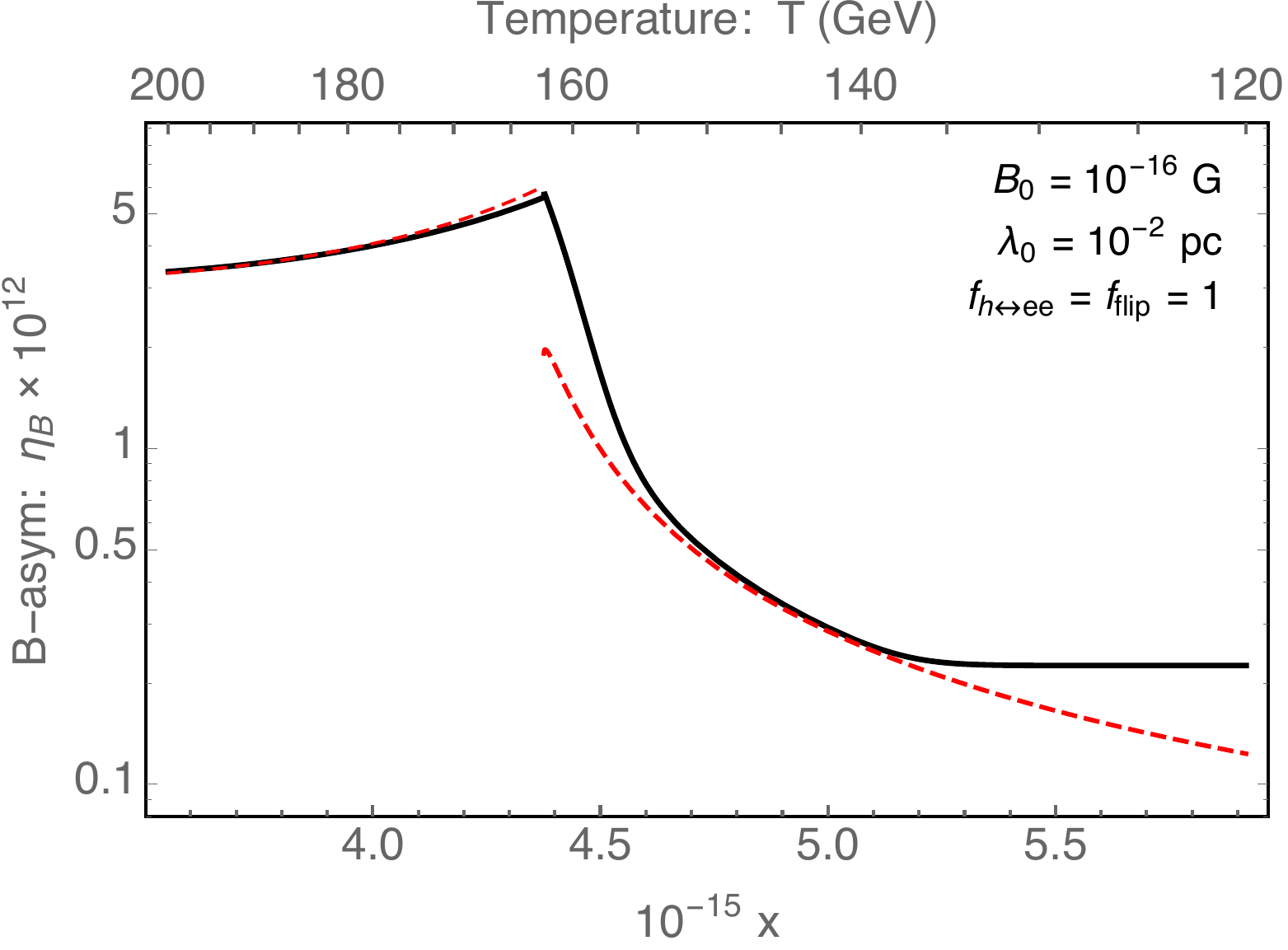} 
\caption{
\label{fig:etaB_versus_x}
Evolution of the baryon asymmetry in the presence of decaying helical magnetic field.  
We take  $B_0 = 10^{-16} \Gauss$ for the field strength today, $\lambda_0 = 10^{-2} \pc$ for the coherence length today, and $f_{\htoee} = f_{\etoe} = 1$ for the spin-flip fudge factor.  The magnetic field is injected at a temperature $T_{\rm ini}$, which ranges from $10^{8}$ to $10^{3} \GeV$ by factors of $10$.  The dashed lines shown the analytic approximations in \erefs{eq:BAUeq_symph}{eq:BAUeq_brkph}.  
}
\end{center}
\end{figure}

Figure~\ref{fig:etaB_versus_x} shows the evolution of the baryon asymmetry $\eta_{B} = n_B / s$ as a function of the dimensionless temporal coordinate $x = M_{0} / T$ with $M_0 \simeq 7.1 \times 10^{17} \GeV$.  
In this plot we take $B_{0} = 10^{-16} \Gauss$, $\lambda_{0} = 10^{-2} \pc$, and $f_{\htoee} = f_{\etoe} = 1$.  
We show different values of the injection temperature $T_{\rm ini}$.  
The baryon asymmetry is observed to rise quickly once the magnetic field is injected.  
This results from the decaying helicity of the right-handed hypermagnetic field, which sources a positive $(B+L)$; if we had taken a left-handed field, the baryon asymmetry would be negative ({\it cf.} \eref{eq:BB_BdB_approx}).  
Note that $(B-L)$ is conserved and vanishes at all times.  
The electroweak sphaleron is unable to washout the $(B+L)$ asymmetry at temperatures $T \gtrsim 10^{4} \GeV$, because the (first generation) electron Yukawa interaction is out of equilibrium, and a lepton asymmetry can be stored in the right-chiral electron $e_R^1$ \cite{Campbell:1992jd}.  
As the temperature decreases to $T \simeq 10^{4} \GeV$ the Yukawa-mediated Higgs-to-electron decay and inverse decay reactions come into equilibrium.  
This would drive the baryon asymmetry to zero exponentially, but the source term, induced by decaying magnetic helicity, softens the decay to a power law \cite{Fujita:2016igl}.  
Once all processes are in equilibrium, the numerical solution matches the approximate analytic solution in \eref{eq:BAUeq_symph} very well.  
In the present example, $\gamma_{\htoee}$ dominates over $\gamma_{\rm y}^{\CME}$ and the baryon asymmetry evolves as $\eta_B \propto x^{-4/3}$.  
The asymptotic solution is insensitive to the initial condition $T_{\rm ini}$, because the equilibrium solution is reached at $T\sim 10^4 \GeV$ when the (first generation) electron Yukawa interaction comes into equilibrium.

As the temperature drops below $T \simeq 162 \GeV$ the electroweak crossover takes place.  
This affects the kinetic equations in three ways.  
First, the hypermagnetic field is converted into an electromagnetic field, which does not source $(B+L)$, see \eref{eq:detaBLdx}. 
Second, the Higgs condensate begins to grow, leading to new sources in the kinetic equations, \eref{eq:source_condensate}, that tend to erase fermion chirality and therefore $(B+L)$.  
Third, the growing weak gauge boson masses leads to a suppression of the $(B+L)$-violating weak sphaleron process, \eref{eq:gammaW_numb}.  
As discussed below \eref{eq:BAUeq_brkph}, the decaying electromagnetic helicity is a source for chirality, which sustains the $(B+L)$ asymmetry from washout by the electroweak sphaleron.
For the parameters chosen in \fref{fig:etaB_versus_x}, the chirality flipping reactions due to the Yukawa interaction overwhelm the chiral magnetic effect, $\gamma_{\htoee}+\gamma_{\etoe}>\gamma_{\rm em}^\CME$.
After the phase transition, the baryon asymmetry is suppressed by a factor of $O(10)$, because of the growing rate of electron chirality-flipping reactions in the presence of the Higgs condensate ($\gamma_{\etoe}$).  
The numerical solution agrees well with our analytic approximation in \eref{eq:BAUeq_brkph} until $T \simeq 135 \GeV$.  
Below this temperature, $(B+L)$ is conserved after the electroweak sphaleron goes out of equilibrium\footnote{In terms of the kinetic equation, this is the temperature at which $1/x \sim \gamma_{w,sph}$.} and the relic baryon asymmetry is well-approximated by \eref{eq:BAUeq_numb_today}.

\begin{figure}[t]
\begin{center}
\includegraphics[width=0.49\textwidth]{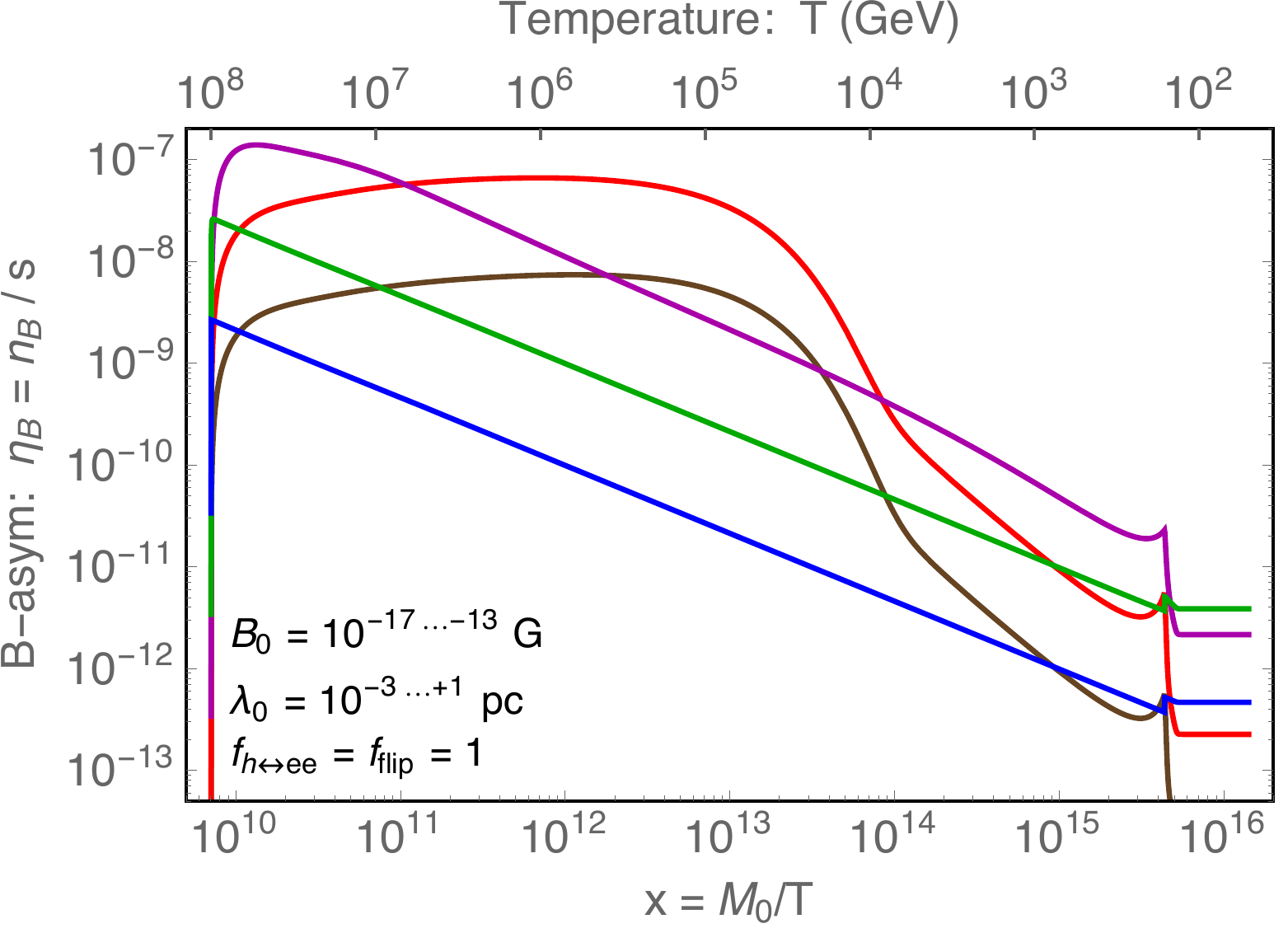} \hfill
\includegraphics[width=0.49\textwidth]{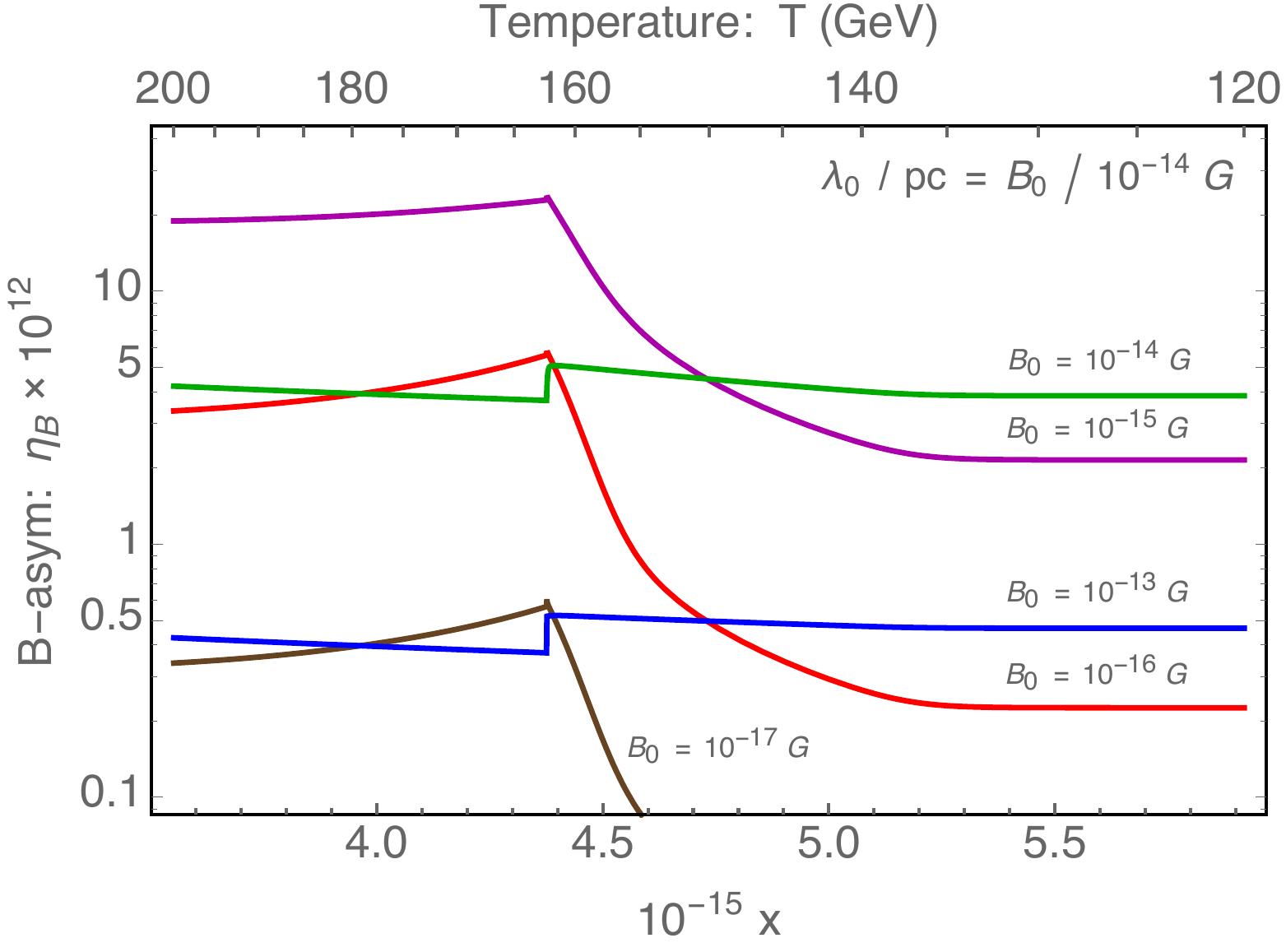} 
\caption{
\label{fig:etaB_versus_B0}
Evolution of the baryon asymmetry for different values of the magnetic field strength today, $B_0$.  The coherence length is given by \eref{eq:lam0_to_B0}, and $f_{\htoee} = f_{\etoe} = 1$ as in \fref{fig:etaB_versus_x}.  As the field strength is varied from $10^{-17} \Gauss$ to $10^{-13} \Gauss$ by factors of $10$ the colors are brown, red, purple, green, and blue.  
}
\end{center}
\end{figure}

Figure~\ref{fig:etaB_versus_B0} reveals how the evolution of the baryon asymmetry $\eta_B$ depends on the relic magnetic field strength $B_0$.  
The coherence length is allowed to vary according to \eref{eq:lam0_to_B0}, which is the expected scaling for a causally generated primordial magnetic field.  
The evolution of $\eta_{B}$ has two qualitatively different behaviors, depending on whether $B_0$ is larger or smaller than about $5 \times 10^{-15} \Gauss$, that can be understood from our analytic solutions in   \erefs{eq:BAUeq_numb_symph}{eq:BAUeq_numb_brkph}.
For a weaker field, washout is limited by the rate of chirality-flipping Yukawa interactions with the Higgs boson and condensate.  
In this case, the chirality-flipping interaction $\gamma_\etoe$ becomes sizeable in the broken phase and results in the slight suppression of the baryon asymmetry.  
For a stronger field, growth of the baryon asymmetry is restricted by chiral magnetic effect.  
(This behavior was not recognized in some previous studies \cite{Anber:2015yca,Fujita:2016igl}.)
In the symmetric phase, the equilibrium solution scales as $\eta_{B} \sim x^{-4/3}$ for weak fields and $x^{-2/3}$ for strong fields, as we showed in \eref{eq:BAUeq_numb_symph}.  

\begin{figure}[t]
\begin{center}
\includegraphics[width=0.49\textwidth]{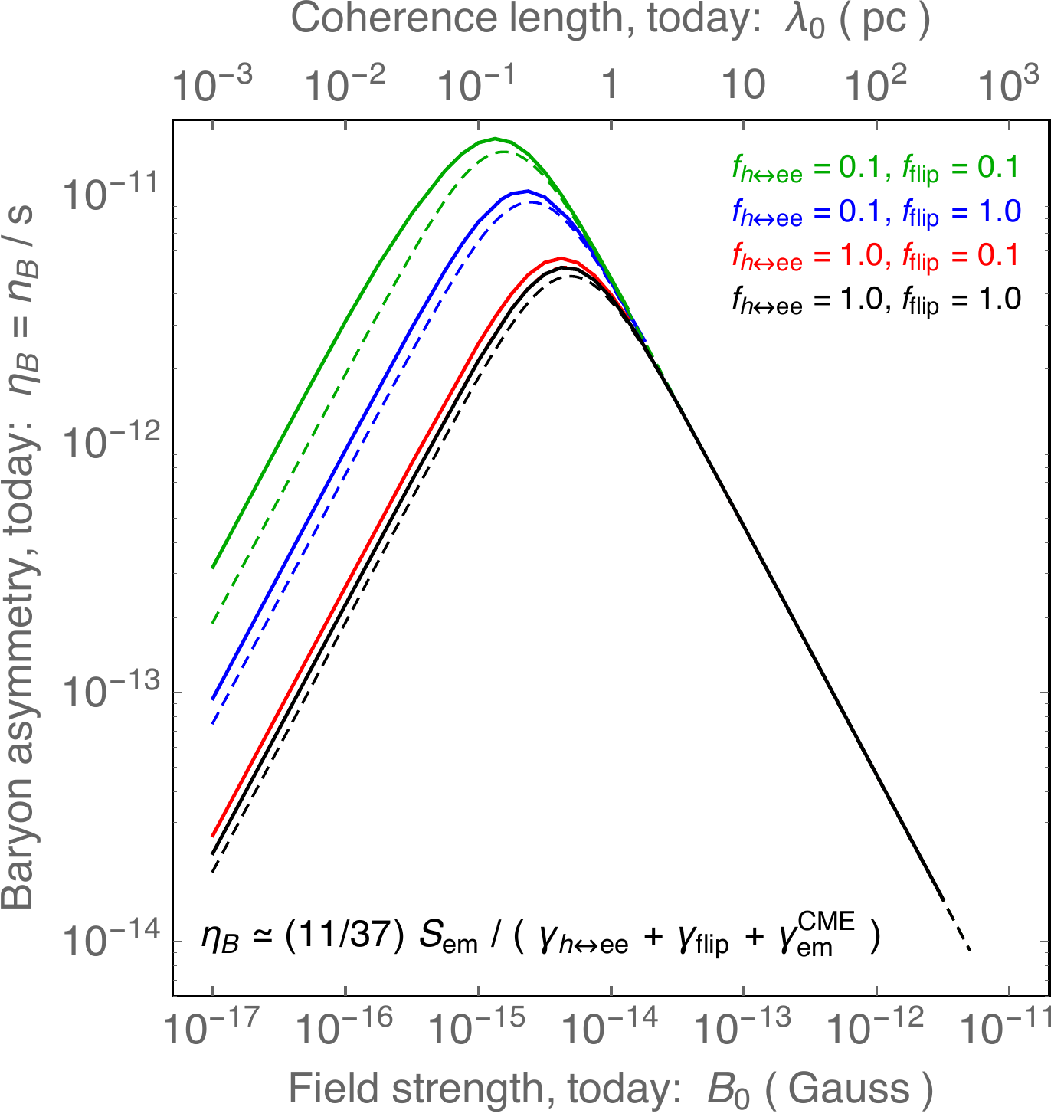} 
\caption{
\label{fig:relicB}
The relic baryon asymmetry as a function of the field strength today $B_0$.  The coherence length $\lambda_0$ satisfies \eref{eq:lam0_to_B0}, and we show a few combinations of the Higgs decay and spin-flip fudge factor, $f_{\htoee}$ and $f_{\etoe}$.  
Solid curves correspond to the numerical solution, and dashed curves indicate the analytic approximation in \eref{eq:BAUeq_numb_today}.  
}
\end{center}
\end{figure}

In \fref{fig:relicB} we show the relic baryon asymmetry $\eta_{B}$ as a function of the magnetic field strength today $B_{0}$ while fixing the coherence length $\lambda_0$ with the relation in \eref{eq:lam0_to_B0}.  
If the field is too weak, the corresponding source term from decaying hypermagnetic helicity is inefficient, and the resulting relic baryon asymmetry is suppressed.  
If the field is too strong, the baryon asymmetry is suppressed instead by the chiral magnetic effect.  
For our best estimates of the electron spin-flip transport coefficients, $f_{\htoee} = f_{\etoe} = 1$, the largest relic baryon asymmetry $\eta_{B} \simeq 5 \times 10^{-12}$ is obtained for $B_0 \simeq 5 \times 10^{-15} \Gauss$.  
This is insufficient to account for the observed baryon asymmetry of the universe, $\eta_{B}^{\rm obs} \simeq 1 \times 10^{-10}$.  
Varying the transport coefficients over a reasonable interval leads to an $O(1)$ change in the relic asymmetry; this indicates the robustness of our result.  

\begin{figure}[t]
\begin{center}
\includegraphics[width=0.49\textwidth]{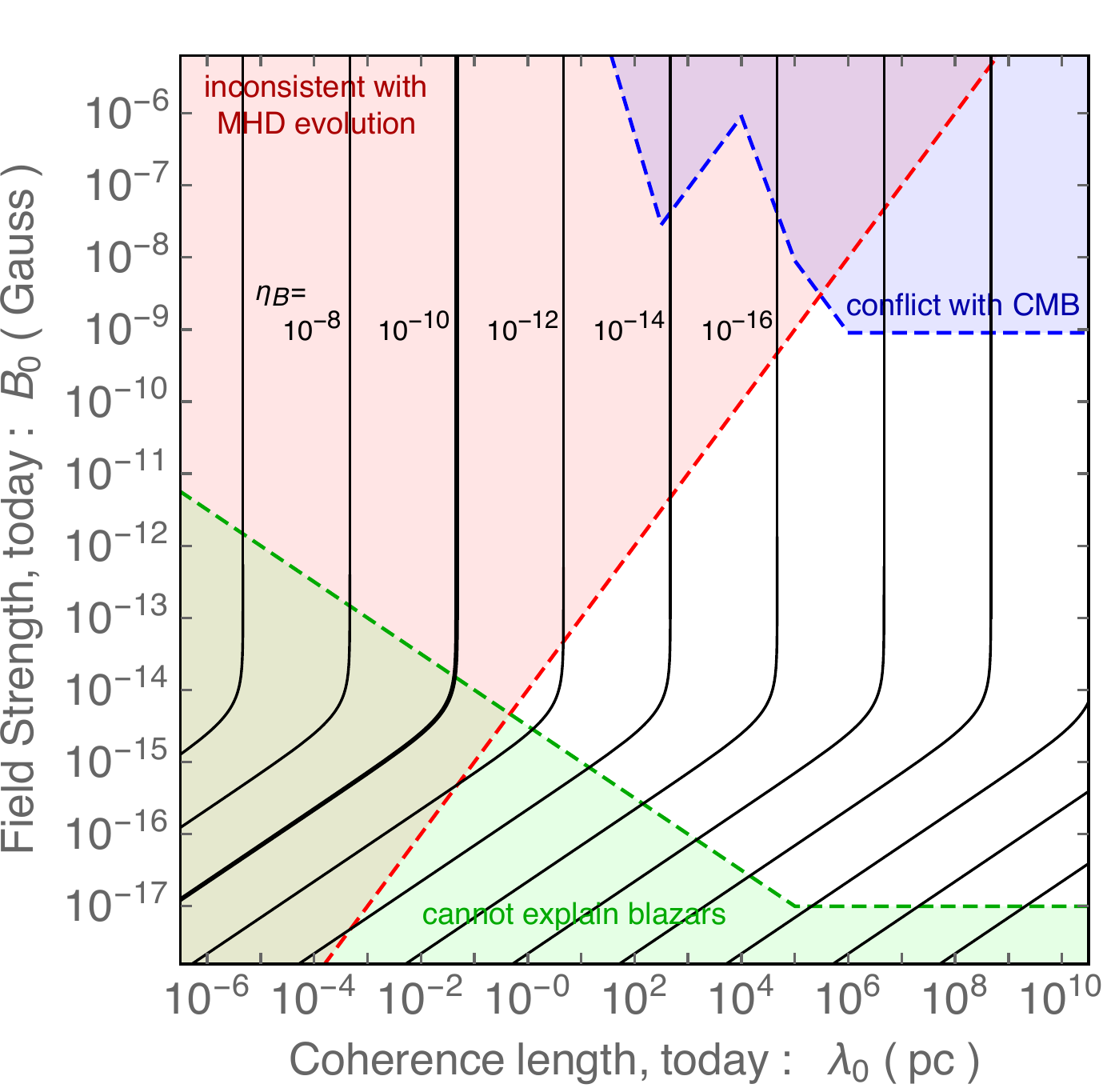} 
\caption{
\label{fig:lam0_B0}
The relic baryon asymmetry over the magnetic field parameter space.  The constraints from CMB, blazars, and MHD are discussed in the text.  We calculate $\eta_B$ from the analytic formula in \eref{eq:BAUeq_numb_today}, which assumes the inverse cascade evolution.  For large correlation length $\lambda_0\gg 1 \pc \, (B_0/10^{-14} \Gauss)$ this calculation is not reliable, since the field instead evolves adiabatically, and the relic baryon asymmetry is further suppressed \cite{Fujita:2016igl}.  
}
\end{center}
\end{figure}

The above results strongly support the validity of our analytic estimate \eref{eq:BAUeq_numb_today}.
Figure \ref{fig:lam0_B0} shows the magnetic field parameter space and predicted baryon asymmetry from the analytic formula \eref{eq:BAUeq_numb_today}.
The constraints are summarized as follows \cite{Durrer:2013pga}.  
On large length scales, a strong field $B_0 \gtrsim 10^{-9} \Gauss$ would induce energy density inhomogeneities at a comparable level to the primordial density perturbations.  
Models falling into the region of parameter space labeled ``conflict with CMB'' are excluded by non-observation of these effects in the cosmic microwave background.  
Measurements of TeV blazar spectra display a deficit of ${\rm GeV}$ photons, which can be explained by a sufficiently strong intergalactic magnetic field that deflects the electromagnetic cascade off the line of sight.  
A weak magnetic field in the region of parameter space labeled ``cannot explain blazars'' cannot accommodate the blazar observations.  
Finally, we have already discussed that a causally-generated primordial field is expected to satisfy $B_0/(10^{-14} \Gauss) \sim \lambda_0/(1\pc)$ today ({\it cf.} \eref{eq:lam0_to_B0}).  
In the region of parameter space labeled ``inconsistent with MHD evolution'', a small scale field transfers its energy to heating the plasma (magnetohydrodynamics) and is not expected to survive until today.  
An arbitrarily large scale magnetic field could be generated acausally during inflation.  
However, a very large scale field will not experience the inverse cascade, and the resultant baryon asymmetry is expected to be smaller than the one we have calculated \cite{Fujita:2016igl}.  

If the relation $\lambda_0/(1\pc) = B_0/(10^{-14} \Gauss)$ is satisfied today ({\it cf.} \eref{eq:lam0_to_B0}), then the relic baryon asymmetry is at most $\eta_B \sim 10^{-11}$, as we saw in \fref{fig:relicB}.  
Upon relaxing this assumption in \fref{fig:lam0_B0}, the observed baryon asymmetry $\eta_B \simeq 10^{-10}$ can be generated for $\lambda_{0} \lesssim 10^{-1} \pc$ and a range of field strengths.
Nominally, this region of parameter space is ``inconsistent with MHD evolution,'' but the boundary defined by \eref{eq:lam0_to_B0} is subject to model-dependent uncertainties. 
For example, \rref{Kahniashvili:2012uj} reports\footnote{In this example, the magnetic field is assumed to arise during the electroweak phase transition and \rref{Kahniashvili:2012uj} uses numerical simulations of magnetohydrodynamics to study its subsequent evolution until recombination.  Since we are interested in models where the magnetic field arises prior to electroweak symmetry breaking, the result of \rref{Kahniashvili:2012uj} is not directly applicable to our calculation.  Nevertheless, we serves to quantify the model-dependent uncertainties behind \eref{eq:lam0_to_B0}.  } a coherence length that is $O(10)$ times smaller than inferred from \eref{eq:lam0_to_B0} for the same field strength.  
Then a maximally helical, primordial magnetic field with strength $B_0 \sim 10^{-14} \Gauss$ and coherence length $\lambda_0 \sim 10^{-1} \pc$ might be responsible for {\it both} the present baryon asymmetry of the Universe {\it and} the observations of blazar spectra.  
It would be interesting to identify specific models of magnetogenesis that are consistent with this ``sweet spot'' in parameter space.

\section{Summary and Discussion}\label{sec:Conclusion}

In this paper, we extend the work of \rref{Fujita:2016igl} to study the generation of a relic baryon asymmetry from the decay of a maximally helical (hyper)magnetic field.  
Due to the Standard Model chiral anomalies, decaying hypermagnetic helicity generates a $(B+L)$ asymmetry. 
Since the asymmetry is continuously generated until the electroweak phase transition, it can arise even with the electroweak sphalerons in equilibrium and $(B-L)=0$.  
Compared to the previous study, we take into account the chiral magnetic effect and the evolution of the system after electroweak symmetry breaking.  

We find that the chiral magnetic effect leads to a suppression of the baryon asymmetry for models with a large magnetic field strength, $B_0 \gtrsim 5 \times 10^{-15} \Gauss$.  
As a result, the baryon overproduction reported in \rref{Fujita:2016igl} is avoided. 
This is understood in the following way \cite{Giovannini:1997eg,Giovannini:1997gp}.  
The chiral magnetic effect causes a magnetic field ${\bm B}$ to induce a current ${\bm j} \propto \mu_5 {\bm B}$ in a medium with a chiral asymmetry $\mu_5$.  
In turn, the current drives an electric field ${\bm E} \propto {\bm j} / \sigma$, and the combination ${\bm E} \cdot {\bm B} \propto \mu_5 B^2 / \sigma$ appears in the kinetic equations due to the chiral anomaly.  
For a strong field, $B = |{\bm B}|$ is large and $\mu_5$ is efficiently erased.  

Considering the evolution soon after the electroweak phase transition, one might expect that the baryon asymmetry is completely washed out.  
For a time, $160 \GeV \gtrsim T \gtrsim 130 \GeV$ the decaying electromagnetic helicity does not source $(B+L)$ and the electroweak sphaleron remains in equilibrium, which tends to erase $(B+L)$.  
Since we take $(B-L)=0$, it would seem that the equilibrium solution corresponds to vanish baryon number, $B=L=0$ \cite{Kuzmin:1985mm}.  
However, our analytic and numerical solutions of the kinetic equations reveal that the baryon asymmetry is not totally washed out.  
Although the decaying electromagnetic field does not source $(B+L)$, it does source a chiral asymmetry; see the discussion below \eref{eq:BAUeq_brkph}.  
Thus, a solution in which all asymmetries are vanishing is not allowed.  
Effectively, a relic baryon asymmetry is maintained by the electromagnetic source term until $T \simeq 135 \GeV$ when the electroweak sphaleron goes out of equilibrium, and afterward baryon number is conserved.  

To further illustrate this point, consider the following toy model.  
Consider the two charge abundances $\eta_L$ and $\eta_R$ which evolve subject to the kinetic equations
\begin{align}
	\frac{d\eta_L}{dx} & = - \gamma_{\rm sph} \eta_L + \gamma_{\rm flip} \bigl( \eta_R - \eta_L \bigr) - \Scal_{\rm em} \\
	\frac{d\eta_R}{dx} & = - \gamma_{\rm flip} \bigl( \eta_R - \eta_L \bigr) + \Scal_{\rm em} 
	\per
\end{align}
As the notation suggests, we can think of $\eta_L$ as the left-chiral baryon number, $\eta_R$ as the right-chiral baryon number, $\gamma_{\rm sph}$ as the electroweak sphaleron transport coefficient, $\gamma_{\rm flip}$ as the Yukawa interaction transport coefficient, and $\Scal_{\rm em}$ as the electromagnetic source term.  
The analog of baryon number, $\eta_L + \eta_R$, is not sourced by $\Scal_{\rm em}$.  
If we had $\Scal_{\rm em} = 0$, the equilibrium solution would be vanishing, $\eta_{L,{\rm eq}} = \eta_{R,{\rm eq}} = 0$.  
Due to the source, the solution is instead ({\it cf.} \eref{eq:BAUeq_brkph})
\begin{align}
	\eta_{L,{\rm eq}} = 0 
	\qquad \text{and} \qquad
	\eta_{R,{\rm eq}} = \frac{\Scal_{\rm em}}{\gamma_{\rm flip}} 
	\per
\end{align}
The sphaleron efficiently washes out the left-chiral baryon number, but the right-chiral baryon number is sustained by the electromagnetic source term.  

The helical (hyper)magnetic field responsible for baryon number generation is expected to persist today in the voids between galaxies and clusters. 
Using the known evolution for a maximally helical magnetic field in a turbulent plasma (inverse cascade), we relate the spectrum of the primordial (hyper)magnetic field to the present day intergalactic magnetic field.  
Thus we calculate the relic baryon asymmetry $\eta_B(t_0)$ in terms of the relic magnetic field strength $B_0$ and coherence length $\lambda_0$.  
For our best estimates of the transport coefficients, \eref{eq:BAUeq_numb_today} gives 
\begin{align}
	\eta_{B} (t_0) 
	\approx \pm (4\times 10^{-12} ) \left( \frac{B_0}{10^{-14} \Gauss} \right)^2 \left( \frac{\lambda_0}{1 \pc} \right)^{-1} \left[ 0.2 + \left( \frac{B_0}{10^{-14} \Gauss} \right)^2 \right]^{-1}
	\com
\end{align}  
where the $+$ is for right-handed helicity and $-$ for left.  
The baryon asymmetry is maximized for $B_0 \sim 5 \times 10^{-15} \Gauss$ and $\lambda_{0} \sim 0.5 \pc$, corresponding to a balance between the two washout processes.  
For larger field strengths, washout induced by the chiral magnetic effect suppresses the baryon asymmetry. 
For smaller field strength, washout is controlled by the (first generation) electron Yukawa interaction.  

Causality arguments suggest that the present day field strength and the correlation length are related by $\lambda_0/(1 \pc)=B_0/(10^{-14}\Gauss)$ ({\it cf.} \eref{eq:lam0_to_B0}).  
With this assumption, the predicted baryon asymmetry is not large enough to explain the observed baryon asymmetry of the Universe, for our best estimates of the transport coefficients.  
We have not calculated the transport coefficients accurately, and this uncertainty can change the prediction. 
If the transport coefficients associated with the electron Yukawa interactions, which lead to washout of $(B+L)$, are smaller by a factor of $10^{-2}$, the present baryon asymmetry can be generated for $B_0\sim 10^{-16\sim17} \Gauss$ and $\lambda_0 \sim 10^{-2\sim3} \pc$.  
While we do not expect that our estimates are so inaccurate, a more careful treatment of the transport equations is desirable. 

Relaxing the relation between $B_0$ and $\lambda_0$, the observed baryon asymmetry is reproduced for $B_{0} \sim 10^{-14} \Gauss$ and $\lambda_{0} \sim 0.1 \pc$; see also \fref{fig:lam0_B0}.  
These parameters are consistent with the causality relation in \eref{eq:lam0_to_B0} up to an $O(10)$ factor, which is within theoretical uncertainties.  
The presence of an intergalactic magnetic field with this spectrum could have a number of interesting implications:  it may have provided the seed field for the galactic dynamo then leading to the micro-Gauss level fields observed in galaxies and clusters, it may help to explain the deficit of secondary GeV photons in blazar observations, and could potentially be probed with future observations of the magnetically broadened cascade halos of TeV blazars.  
Therefore, it would be interesting to identify specific models of magnetogenesis that are consistent with this ``sweet spot'' in parameter space.

\quad \\
\noindent
{\bf Acknowledgments:} 
KK~acknowledges support from the DOE for this work under Grant No. DE-SC0013605.
AJL is supported at the University of Chicago by the Kavli Institute for Cosmological Physics through grant NSF PHY-1125897 and an endowment from the Kavli Foundation and its founder Fred Kavli.  
We thank Daniel Chung and Petar Pavlovic for discussing the calculation of transport coefficients, and we thank Tomohiro Fujita and Tanmay Vachaspati for comments on the draft. 

\appendix

\section{Calculation of Transport Coefficients}\label{app:transport_coeff}

In this appendix we provide additional details for the calculation of transport coefficients.  
Here we rely on the formalism based on the Schwinger-Dyson equations studied in \rref{Cirigliano:2006wh} (see also Refs.~\cite{Chung:2009qs,Riotto:1998zb,Lee:2004we}).  
We do not discuss the transport coefficients for the weak gauge interactions explicitly, but the calculation is similar to the Yukawa interactions discussed below.  

\subsection*{Decay and Inverse Decay Mediated by Yukawa Interaction}\label{app:Yukawa}

The Yukawa interactions mediate decay and inverse decay reactions, which are listed in \eref{eq:Yukawa_reactions}.  
For the leptons, the corresponding transport coefficients are \cite{Cirigliano:2006wh} (see also \rref{Chung:2009qs})
\begin{align}\label{eq:Gam_to_IF}
	\gamma_{\Nhe}^{ij} & = \frac{12 |y_{e}^{ij}|^2}{T^3} \Ical_{F}(m_{e_R^j}, m_{\nu_L^i}, m_{\phi^+}) \\
	\gamma_{\Ehe}^{ij} & = \frac{12 |y_{e}^{ij}|^2}{T^3} \Ical_{F}(m_{e_R^j}, m_{e_L^i}, m_{\phi^0}) 
\end{align}
where $y_{e}^{ij}$ is the electron Yukawa matrix, and the arguments of $\Ical_{F}$ are the respective particle masses.  
Analogous expressions for the quarks are obtained by including a color factor $N_{\rm c}=3$ and an obvious change of labels.  
The kinematic function is given by 
\begin{align}\label{eq:IF_def}
	& \Ical_{F}(m_{1},m_{2},m_{\phi}) = \frac{1}{16\pi^3 T} \bigl( m_{1}^2 + m_{2}^2 - m_{\phi}^2 \bigr) \int_{m_1}^{\infty} \! \ud \omega_{1} \, \int_{\omega_{\phi}^{-}}^{\omega_{\phi}^{+}} \! \ud \omega_{\phi} 
	\nn & \quad 
	\times \Bigl\{ 
	n_{B}(\omega_{\phi}) \bigl[ 1 - n_F(\omega_1) \bigr] n_{F}(\omega_1 - \omega_{\phi}) \Theta(m_1 - m_2 - m_{\phi}) 
	\nn & \quad \qquad
	- n_{B}(\omega_{\phi}) \bigl[ 1 - n_F(\omega_1) \bigr] n_{F}(\omega_1 - \omega_{\phi}) \Theta(m_{\phi} - m_1 - m_2) 
	\nn & \quad \qquad
	+ n_{B}(\omega_{\phi}) n_{F}(\omega_{1}) \bigl[ 1 - n_{F}(\omega_{1} + \omega_{\phi}) \bigr] \Theta(m_2 - m_1 - m_{\phi}) \Bigr\}
\end{align}
where 
\begin{align}
	n_{B}(\omega) = \frac{1}{e^{\omega/T} - 1}
	\qquad \text{and} \qquad
	n_{F}(\omega) = \frac{1}{e^{\omega/T} + 1}
\end{align}
are the Bose-Einstein and Fermi-Dirac phase space distribution functions, and 
\begin{align}
	\omega_{\phi}^{\pm} = \frac{1}{2m_1^2} \Bigl\{ 
	& \omega_{1} \, \bigl| m_{\phi}^2 + m_{1}^2 - m_{2}^2 \bigr| 
	\pm \sqrt{ (\omega_{1}^2 - m_{1}^2) \bigl( m_{1}^2 - (m_{2}+m_{\phi})^2\bigr) \bigl( m_{1}^2 - (m_{2} - m_{\phi})^2 \bigr) } \Bigr\}
\end{align}
are the limits of integration.  

In \eref{eq:IF_def}, the step functions, $\Theta(m)$, ensure that the corresponding decay or inverse decay channel is kinematically accessible.  
If the three-body reactions are kinematically blocked, the transport coefficient can arise from 2-to-2 scattering.  
Since these contributions are generally suppressed by an additional factor of coupling-squared, we expect the three-body channels to dominate when they are open.  

At high temperature in the symmetric phase, the thermal masses are given by \cite{Joyce:1994zn}
\begin{align}
	m_{\nu_L^i}^2 = m_{e_L^i}^2 & = \Bigl( \frac{3\pi}{8} \alpha_{\rm w} + \frac{\pi}{2} y_{L_L}^2 \alpha_{\rm y} \Bigr) T^2 \simeq (0.207 \, T)^2 \\
	m_{e_R^i}^2 & = \Bigl( \frac{\pi}{2} y_{e_R}^2 \alpha_{\rm y} \Bigr) T^2 \simeq (0.123 \, T)^2 \\
	m_{\phi^+}^2 = m_{\phi^0}^2 & = 2 \Bigl( \frac{3\pi}{8} \alpha_{\rm w} + \frac{\pi}{2} y_{\Phi}^2 \alpha_{\rm y} + \frac{y_t^2}{8} + \frac{\lambda}{8} \Bigr) T^2 \simeq (0.602 \, T)^2 
	\per
\end{align}
Thus the kinematically accessible channels are Higgs decay and inverse decay ($m_{\phi} > m_{1} + m_{2}$).  
The transport coefficients can be written as 
\begin{align}
	\gamma_{\Nhe}^{ij} & = \frac{|y_{e}^{ij}|^2}{8\pi} \frac{m_{\phi^+}^2 - m_{e_R^j}^2 - m_{\nu_L^i}^2}{T^2} \, \bar{\Ical}_{F}(m_{e_R^j}, m_{\nu_L^i}, m_{\phi^+}) \\
	\gamma_{\Ehe}^{ij} & = \frac{|y_{e}^{ij}|^2}{8\pi} \frac{m_{\phi^0}^2 - m_{e_R^j}^2 - m_{e_L^i}^2}{T^2} \, \bar{\Ical}_{F}(m_{e_R^j}, m_{e_L^i}, m_{\phi^0})
\end{align}
where the rescaled, dimensionless integral is 
\begin{align}\label{eq:IFbar_def}
	\bar{\Ical}_{F} & = \frac{6}{\pi^2 T^2} \int_{m_1}^{\infty} \! \ud \omega_{1} \, \int_{\omega_{\phi}^{-}}^{\omega_{\phi}^{+}} \! \ud \omega_{\phi} \, n_{B}(\omega_{\phi}) \bigl[ 1 - n_F(\omega_1) \bigr] n_{F}(\omega_1 - \omega_{\phi}) 
	\per
\end{align}
Using the thermal masses above, we evaluate the integral numerically to obtain $\bar{\Ical}_{F} \simeq 0.250$ in both $\gamma_{\Nhe}^{ij}$ and $\gamma_{\Ehe}^{ij}$.  
The temperature dependence cancels out in the symmetric phase, since it is the only energy scale.  

To obtain an analytic expression for $\bar{\Ical}_F$ we focus on the regime, $m_{1} , m_{2} \ll m_{\phi} \ll T$.  
Evaluating \eref{eq:IFbar_def} gives $\bar{\Ical}_F \approx (6 \ln 2) / \pi^2 \simeq 0.421$, and the transport coefficients become
\begin{align}
	\gamma_{\Nhe}^{ij} \approx \gamma_{\Ehe}^{ij} \approx \frac{6 \ln 2}{\pi^2} \frac{|y_{e}^{ij}|^2}{8\pi} \frac{m_{h}^2}{T^2}
	\per
\end{align}
This agrees with the result of \rref{Campbell:1992jd}, where the first factor is $2(\ln 2)^2 / (3 \zeta(3)) \simeq 0.266$ instead of $(6 \ln 2) / \pi^2$.  

\subsection*{Spin-Flip Mediated by Yukawa Interaction}\label{app:fermion_cond}

As we discussed in \eref{eq:Hcond_reactions}, fermions experience a spin-flip interaction by scattering with the Higgs condensate.  
For the leptons, the corresponding transport coefficient is written as \cite{Lee:2004we}
\begin{align}
	\gamma_{\Ee}^{ij} = \frac{6}{\pi^2} |y_{e}^{ij}|^2 \frac{v(T)^2}{T^2} \Ical(T)
	\com
\end{align}
and for the quarks there is an additional color factor ($N_{\rm c} = 3$).  
The momentum integral is 
\begin{align}
	\Ical(T) = \frac{1}{T^2} \int_{0}^{\infty} \! k^2 \ud k \, {\rm Im} \Biggl\{ 
	& \frac{Z_p^L(k) Z_p^R(k)}{\Ecal_p^L + \Ecal_p^R} \Bigl[ h_F(\Ecal_p^L) + h_F(\Ecal_p^R) \Bigr] \nn
	& \quad + \frac{Z_p^L(k) Z_h^R(k)^{\ast}}{\Ecal_p^L - \Ecal_h^{R \ast}} \Bigl[ h_F(\Ecal_p^L) + h_F(\Ecal_h^{R\ast}) \Bigr] 
	+ \bigl( p \leftrightarrow h \bigr)
	\Biggr\}
	\per
\end{align}
The temperature $T$ enters explicitly through the fermionic thermal function 
\begin{align}
	h_F(\Ecal) = \frac{e^{\Ecal/T}}{\bigl( e^{\Ecal/T} + 1 \bigr)^2}
	\per
\end{align}
The self-energy has poles in the complex plane at 
\begin{align}
	\Ecal_{p,h}^{L,R}(k) = \omega_{p,h}^{L,R}(k) - i \Gamma_{p,h}^{L,R}(k)
\end{align}
with $\omega$ and $\Gamma$ real.  
These corresponds to particle ($p$) and holes ($h$) of left ($L$) and right ($R$) handed chirality.  
The corresponding residues are denoted by $Z_{p,h}^{L,R}(k)$.  
Without loss of generality we can write
\begin{align}
	\omega_{p,h}^{L,R}(k)^2 = k^2 + M_{p,h}^{L,R}(k)^2 
\end{align}
where $M^2$ will be momentum dependent if the dispersion relation is non-trivial.  

To evaluate $\Ical$ we make the following simplifying assumptions.  We assume (1) that the residues are real, (2) that the masses, widths, and residues are independent of momentum $k$, (3) that the widths and masses are hierarchical $\Gamma \sim g^2 T$ and $M \sim gT$ with $g \ll 1$, and (4) that the hole contributions are negligible ($Z_h^{L,R} \approx 0$).  
Under these assumptions, the integral is found to be 
\begin{align}\label{eq:IF_approx}
	\Ical_F \approx Z_p^L Z_p^R \frac{\Gamma_{p}^{L} + \Gamma_{p}^{R}}{2T}
	\per
\end{align}
Sub-leading corrections are suppressed by $M^2 / T^2 \sim g^2 \ll 1$.  
Taking the residues to be $O(1)$ numbers, the transport coefficient is estimated as 
\begin{align}
	\gamma_{\Ee}^{ij} \approx \frac{6}{\pi^2} |y_{e}^{ij}|^2 \frac{v(T)^2}{T^2} \frac{\Gamma}{T}
\end{align}
where $\Gamma = (\Gamma_p^L + \Gamma_p^R)/2$ is the average thermal width.  
This approximation motivates our estimates of the transport coefficients in \eref{eq:gammaYukM_approx}.  

\section{Thermal Higgs Mass and Condensate}\label{app:thermal_mass}

The SM crossover has been studied on the lattice in \rref{DOnofrio:2015mpa}.  
By fitting an analytic function to the numerical lattice result (Figure 3 of \rref{DOnofrio:2015mpa}), we determine an empirical formula for the growth of the condensate soon after the phase transition: 
\begin{align}\label{eq:vT_empirical}
	v(T) \simeq 
	\begin{cases}
	0 & \quad , \ T > 162 \GeV \\
	0.23 T \sqrt{162 - T/{\rm GeV}} & \quad , \ T < 162 \GeV 
	\end{cases}
	\per
\end{align}
This formula matches the lattice result very well in the temperature range $140 \GeV < T < 162 \GeV$.  
When the temperature becomes very low, $T \lesssim 110 \GeV$, this formula implies that $v(T)$ will start to decrease, which is not the correct behavior.  
Thus we focus our analysis on temperatures $T > 130 \GeV$ where we expect the empirical fitting formula to be reliable.  

\begin{figure}[t]
\hspace{0pt}
\vspace{-0in}
\begin{center}
\includegraphics[width=0.4\textwidth]{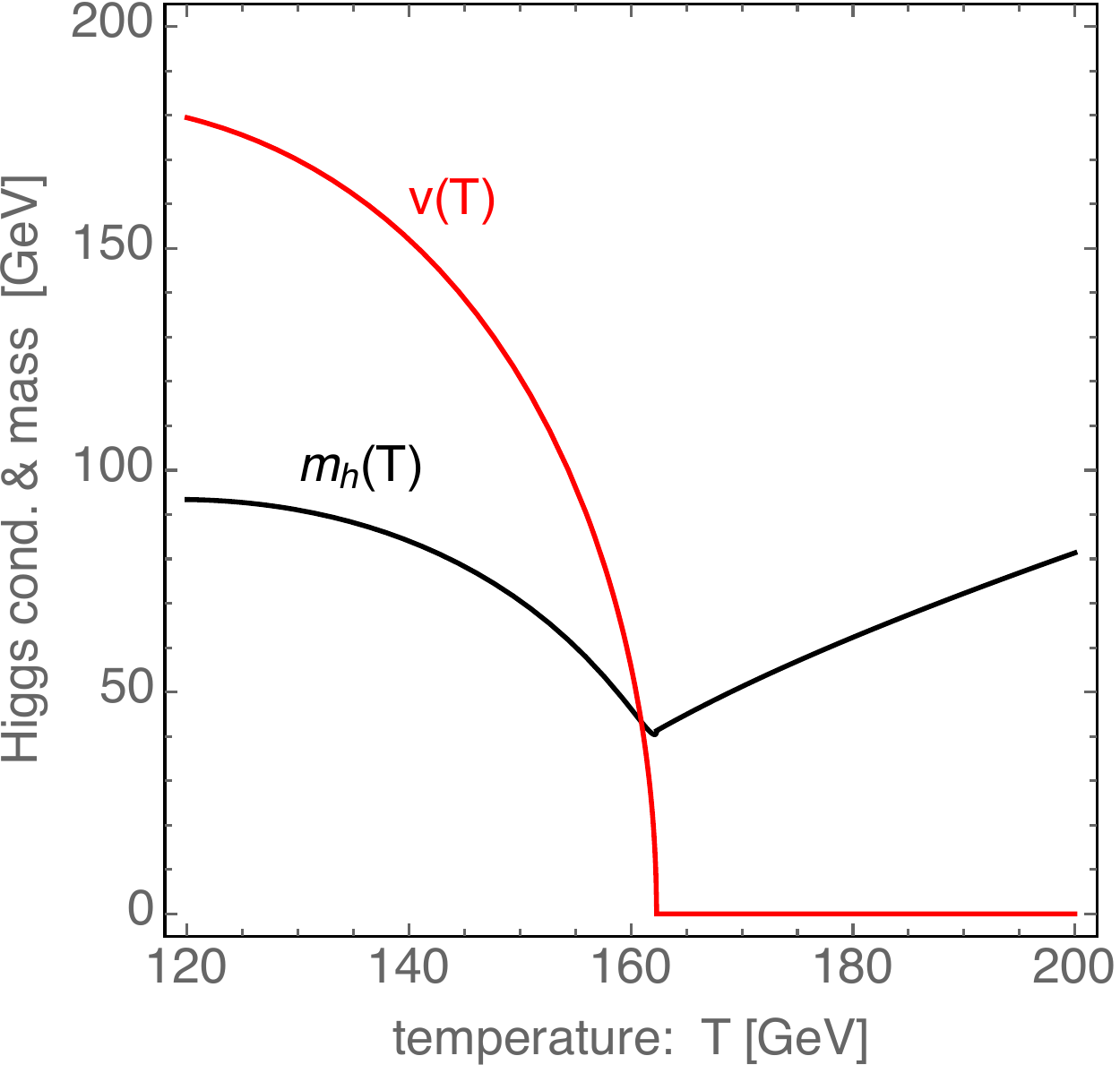}
\caption{
\label{fig:Higgs_versus_temp}
The Higgs condensate $v(T)$ from \eref{eq:vT_empirical} and thermal Higgs mass $m_h(T)$ from \eref{eq:Higgs_therm_mass}.  
}
\end{center}
\end{figure}

During the electroweak phase transition, particle masses are affected by the growth of the Higgs condensate.  
We identify the physical Higgs field $h(x)$ by writing $\phi^0 = \bigl( v(T) + h \bigr) / \sqrt{2}$ where $v(T)$ is the value of the Higgs condensate.  
During the phase transition, $v(T)$ is given by the empirical fitting formula in \eref{eq:vT_empirical}.  
The thermal Higgs boson mass is \cite{Quiros:1999jp}
\begin{align}\label{eq:Higgs_therm_mass}
	m_{h}^2(T) & = 2 D \bigl( T^2 - T_0^2 \bigr) - 6 E T v(T) + 3 \lambda v(T)^2 
\end{align}
where
\begin{align}
	D & = \frac{1}{8v^2} \left( 2 m_W^2 + m_Z^2 + 2m_t^2 + \frac{1}{2} m_h^2 \right) \simeq 0.18 \\
	T_0^2 & = \frac{m_h^2}{4D} \simeq (147 \GeV)^2 \\
	E & = \frac{2 m_W^3 + m_Z^3}{4 \pi v^3} \simeq 0.0096
	\per
\end{align}
In the symmetric phase, $m_h^2(T)/T^2 \approx 2D \simeq 0.36$.  
The condensate and thermal Higgs mass are shown in \fref{fig:Higgs_versus_temp}.  

\section{Relating Primordial and Present Magnetic Field Spectra}\label{app:Bplamb00}

Let $B_0$ and $\lambda_0$ be the mean field strength and coherence length of the relic magnetic field today, and let $B_p(t)$ and $\lambda_B(t)$ correspond to these values in the early universe at time $t$.  
We write the comoving field strength and coherence length as $\tilde{B}(\tau) = a(t)^2 B_p(t)$ and $\tilde{\lambda}(\tau) = a(t)^{-1} \lambda_B(t)$ where $a(t)$ is the FRW scale factor and $\tau$ is the conformal time variable.  
Between recombination and today, the plasma is neutral and, to a good approximation, the magnetic field evolves adiabatically.  
During adiabatic evolution, the comoving field strength and coherence length are conserved, and we can write 
\begin{align}
	\tilde{B}_{\rm rec} = B_0
	\qquad \text{and} \qquad
	\tilde{\lambda}_{\rm rec} = \lambda_0
	\per
\end{align}
Prior to recombination, the magnetic field freely decays in the turbulent plasma.  
We estimate the coherence length $\lambda_B(t)$ as the largest eddy size that has been processed by time $t$.  
This implies the relation $\tilde{\lambda}(\tau) = C v_A(\tau) \tau$ where $C$ is a constant of proportionality, and $v_A \propto \tilde{B}$ is the Alfv\'en velocity.  
Turbulence continues until recombination,\footnote{At late times, the turbulent evolution is disrupted by intermittent periods of viscous damping (as the neutrinos and photons decouple) during which the inverse cascade scaling is halted.  However, the overall behavior is roughly consistent with the inverse cascade scaling law.  For a discussion of these details, see \rref{Banerjee:2004df}.  } and we can write 
\begin{align}
	\frac{\tilde{\lambda}(\tau)}{\tilde{B}(\tau) \tau} = \frac{\tilde{\lambda}_{\rm rec}}{\tilde{B}_{\rm rec} \tau_{\rm rec}}
	\per
\end{align}
We also assume that the magnetic field is maximally helical; then the comoving helicity density is estimated as $\pm \tilde{\lambda}(\tau) \tilde{B}(\tau)^2$.  
To a good approximation helicity is conserved,\footnote{In our calculation of anomalous baryon-number generation, it is important that the magnetic helicity is {\it not conserved} ({\it cf.} \eref{eq:Bdot}).  However, for the parameters of interest, the magnetic helicity decays slowly enough that we can treat it as a conserved quantity for the purposes of relating magnetic spectra in the early universe and today \cite{Fujita:2016igl}.  In other words, we could include an $O(1)$ suppression factor in \eref{eq:heli_conserved}, but this would not affect our results significantly.  } and we can write 
\begin{align}\label{eq:heli_conserved}
	\tilde{\lambda} \tilde{B}^2 = \tilde{\lambda}_{\rm rec} \tilde{B}_{\rm rec}^2
	\per
\end{align}
Solving these equations gives the field strength and coherence length, 
\begin{align}\label{eq:Blam_from_B0lam0}
	B_p = \left( \frac{a}{a_0} \right)^{-2} \left( \frac{\tau}{\tau_{\rm rec}} \right)^{-1/3} B_0
	\qquad \text{and} \qquad	
	\lambda_B = \left( \frac{a}{a_0} \right) \left( \frac{\tau}{\tau_{\rm rec}} \right)^{2/3} \lambda_0
	\per
\end{align}
These relations display the characteristic (inverse cascade) scaling behavior of freely decaying helical magnetic fields.  
Upon expressing $a(t)$ and $\tau$ in terms of temperature $T$, we obtain \erefs{eq:Bp_numb}{eq:lamB_numb}.  

These relations were recently derived in \rref{Fujita:2016igl}, which found a different scaling behavior: $B_p \propto \lambda_0^{1/3} B_0^{2/3}$ and $\lambda_B \propto \lambda_0^{1/3} B_0^{2/3}$.  
That result is obtained by combining $\lambda_B = C B_p t$ with $\lambda_B B_p^2 \propto \lambda_0 B_0^2$ to arrive at $B_p \propto \bigl( \lambda_0 B_0^2 / C \bigr)^{1/3}$ and $\lambda_B \propto \bigl( C^2 \lambda_0 B_0^2 \bigr)^{1/3}$.  
Taking the constant of proportionality as $C=1$ gives the scaling in \rref{Fujita:2016igl}.  
Taking instead $C \propto \lambda_0 / B_0$ gives the scaling in \eref{eq:Blam_from_B0lam0}.  
Implicitly, the derivation of \eref{eq:Blam_from_B0lam0} assumes that the back reaction from the chiral magnetic effect is not so strong as to disrupt the inverse cascade scaling relation \cite{Fujita:2016igl}.

\bibliographystyle{h-physrev5}
\bibliography{BAU_from_MagHeli}

\end{document}